\documentclass[article]{IEEEtran}

\usepackage{epsf}
\usepackage{graphicx}
\usepackage{epsfig,latexsym,amsmath,epsf,amssymb,amsfonts}
\usepackage{amssymb,cite}
\usepackage{placeins}
\usepackage{epsfig}
\usepackage{epstopdf}
\usepackage{amsmath}
\usepackage{subcaption}
\usepackage{placeins}
\usepackage{dsfont}
\usepackage{stackrel}
\usepackage{yfonts,color}
\usepackage{lettrine}
\usepackage{caption}
\usepackage{mathtools}
\mathtoolsset{showonlyrefs}
\usepackage{epstopdf}
\usepackage{tikz}
\usetikzlibrary{arrows}
\usetikzlibrary{plotmarks}
\usepackage{cite}
\usepackage{color}
\usepackage{standalone}
\usepackage{pstricks}
\usepackage{graphicx}
\usepackage{amsmath}
\usepackage{amssymb}
\usepackage{mathtools}
\usepackage{psfrag}
\usepackage[ulem=normalem]{changes}
\usepackage{tikz,pgfplots}

\usepackage{verbatim}
\usepackage{algorithm}
\usepackage{algpseudocode}

\usepackage{bm}
\usepackage{fixltx2e}
\usepackage{calc, pgf, xcolor}
\usepackage{colordvi,psfrag}
\usepackage[font=small,skip=0pt]{caption}

\usepackage{nicefrac}

\newtheorem{theorem}{Theorem}

\newtheorem{lemma}{Lemma}

\newtheorem{definition}{Definition}

\newtheorem{remark}{Remark}

\algnewcommand\Or{\textbf{or}}

\newcommand\scalemath[2]{\scalebox{#1}{\mbox{\ensuremath{\displaystyle #2}}}}

\newcommand{\C}{C_{\text{sym}}}

\newcommand{\dc}{ \Delta C}
\newcommand{\dr}{ \Delta R}

\newcommand{\T}{\dagger}
\newcommand{\lth}{\ell^\text{th}}
\newcommand{\kth}{k^\text{th}}
\newcommand{\mth}{m^\text{th}}

\def\argmin{\operatornamewithlimits{arg\,min}}

\IEEEoverridecommandlockouts
\allowdisplaybreaks
\makeatletter
\newcommand*{\rom}[1]{\expandafter\@slowromancap\romannumeral #1@}
\makeatother

\begin{document}

\title {Integer-Forcing Architectures for\\ Uplink Cloud Radio Access Networks}
\author{\IEEEauthorblockN{Islam El Bakoury and Bobak Nazer,}
\thanks{This work was supported by NSF CAREER grant CCF-1253918. The material in this paper was presented in part at the 2017 Allerton Conference on Communications, Control and Computing, Monticello, IL~\cite{bn17}.}
\thanks{I. El Bakoury is with Qualcomm, Boxborough, MA.}\thanks{B. Nazer is with the Department of Electrical and Computer Engineering, Boston University, Boston, MA 02215 USA (e-mail: bobak@bu.edu).}
}

\maketitle

\begin{abstract} Consider an uplink cloud radio access network where users are observed simultaneously by several base stations, each with a rate-limited link to a central processor, which wishes to decode all transmitted messages. Recent efforts have demonstrated the advantages of compression-based strategies
that send quantized channel observations to the central processor, rather than attempt local decoding. We study the setting where channel state information is not available at the transmitters, but known fully or partially at the base stations. We propose an end-to-end integer-forcing framework for compression-based uplink cloud radio access, and show that it operates within a constant gap from the optimal outage probability if channel state information is fully available at the base stations. We demonstrate via simulations that our framework is competitive with state-of-the-art Wyner-Ziv-based strategies. 
\end{abstract}

\section{Introduction}
Cloud radio access networks (C-RANs) have emerged as a promising framework for next-generation wireless communication systems~\cite{cran_book,smpsy16}, since they have the potential to reduce the decoding complexity, energy consumption, and interference caused by the growing density of mobile devices~\cite{peng2015system,li2017green} as well as the sharply-increasing demand for higher data rates~\cite{cisco}. The basic architecture of a C-RAN consists of many users that communicate to several basestations (BSs) over a shared wireless channel. Each BS has a finite-capacity fronthaul link to a central processor (CP). For uplink communication, the BSs send functions of their observations to the C-RAN, which employs a joint decoding strategy to recover the users' messages. Conversely, for downlink communication, the CP utilizes a joint encoding strategy to generate the signals to be sent by the BSs, which are then sent over the finite-capacity fronthaul, emitted on the wireless channel by the BSs, and finally decoded by the individual users. In this paper, we focus exclusively on the uplink C-RAN scenario.

Since an uplink C-RAN can be viewed as a particular instance of a two-hop relay network, we can design a transceiver architecture by drawing upon powerful relaying strategies such as decode-and-forward\cite{ltw04,nhh04,ce79,kgg05}, compress-and-forward\cite{ce79,kgg05,lkec11} and, more recently, compute-and-forward\cite{ng11IT,nw12,hc13}. Recall, that, in the decode-and-forward strategy, the relays recover the individual codewords, in the compute-and-forward strategy, they recover integer-linear combinations of the codewords, and, in the compress-and-forward strategy, they quantize their analog observations. In our C-RAN context, decode-and-forward amounts to each BS recovering one or more codewords while treating the rest as noise, and does not benefit from the joint processing power of the CP. Compute-and-forward can attain higher rates by decoding integer-linear combinations that closely match the channel realization at each BS, and sending these linear combinations to the CP to solve for the users' messages. In contrast, compress-and-forward offloads most of the decoding process to the CP, which has access to the quantized channel output from each BS. This transforms the challenging relaying problem into a virtual multiple-access channel (MAC), and enables us to employ ``off-the-shelf'' equalization and decoding techniques. While compute-and-forward can outperform compress-and-forward for certain channel realizations~\cite{nsgs09}, the ensemble average performance of compress-and-forward is superior, especially for moderate-to-high fronthaul rates. Furthermore, the BSs do not need to know the users' codebooks, and can instead just forward their quantized channel observations to the CP~\cite{SSVS09} in an oblivious fashion. Finally, from a theoretical perspective, compress-and-forward can be used to establish a ``constant-gap'' capacity approximation for C-RANs~\cite{zy14}. For these reasons, we focus exclusively on compression-based C-RAN architectures in this paper.

Clearly, the highest possible rates can be attained by employing simultaneous (or joint) decoding for both the decompression and channel decoding stages. However, the implementation complexity of simultaneous decoding scales exponentially with the number of users, and thus there has been significant interest in developing low-complexity architectures with comparable performance. In particular, a series of recent papers~\cite{zy13,psss13,psss14,zy14,YW16,ZXYC16} have proposed a sequential decoding architecture based on Wyner-Ziv (WZ) compression~\cite{wz76} as well as optimization algorithms for the associated parameters. Through extensive simulations, this line of work has demonstrated the performance advantages of WZ compression over ``single-user'' compression strategies that ignore the correlations between the BSs' observations. Moreover, if the BSs and CP have perfect channel state information (CSI) and the users have enough information to set their rates, then it can be shown that WZ compression combined with successive interference cancellation (SIC) channel decoding can attain the same sum rate as simultaneous source and channel decoding, and operates within a constant gap of the sum capacity~\cite{ZXYC16}. Thus, sequential compression and channel decoding architectures are a compelling framework for C-RAN implementations.

In this paper, we introduce an alternative uplink C-RAN architecture based on \textit{integer-forcing} (IF) source~\cite{oe17} and channel coding~\cite{zneg14}, and demonstrate that it can sometimes outperform the sequential decoding architectures mentioned above. The underlying principle of integer-forcing is that lattice codebooks are closed under addition, and therefore \textit{any integer-linear combination of lattice codewords is itself a codeword.} Recall that, for single-user compression, the quantization rate to attain a fixed distortion depends on the source variance. For integer-forcing source coding, the rate depends instead on the variance of the selected integer-linear combination of the sources. Specifically, the BSs employ (oblivious) lattice quantization and the CP takes integer-linear combinations of its received codewords to minimize the effective source variance prior to reconstruction, and only afterwards solves the resulting system of linear equations for the desired quantized sources. Integer-forcing channel coding is quite similar: the CP first decodes integer-linear combinations of the codewords, chosen to minimize the noise variance, and afterwards solves for the transmitted codewords, thus revealing the users' messages. We join these two integer-forcing strategies end-to-end as the basis for our integer-forcing uplink C-RAN architecture. For the important special case of symmetric rates, this architecture admits a simple realization with a single quantization codebook shared by all the BSs and a single channel codebook shared by all the users, which can be nearly parallelized, except for the final step to solve the linear equations. A more sophisticated variation employs a different codebook at each BS, thus allowing us to tune the rates based on CSI as well as benefit from sequential decoding. 

 The primary motivation for this paper is to evaluate the performance of integer-forcing for uplink C-RANs and compare it to that of WZ strategies. Although prior IF work has explored its performance for source coding and channel coding separately, they must be examined jointly in the context of compression-based C-RANs. One of our main contributions are derivations of the end-to-end rate expressions for integer-forcing, along with algorithms for selecting the associated parameters. In particular, we explore both parallel and sequential decoding architectures for integer-forcing source coding. 

We evaluate the performance of WZ and IF architectures via numerical simulations and demonstrate that, in certain regimes, IF can outperform WZ. Specifically, we consider scenarios where there is no CSI at the transmitters (CSIT) and CSI at the receivers (CSIR) is either global (i.e., available to both BSs and CP) or local (i.e., fully available to the CP but each BS only knows its own channel coefficients). For the latter scenario, we propose an opportunistic IF strategy that helps to mitigate the performance loss due to the lack of global CSIR. The advantage of IF can be partially attributed to the fact that it performs well when the rates are symmetric, whereas WZ attains the highest sum rates when the individual rates are set to attain a corner point of the capacity region (which would require CSIT). 

To complement our numerical studies, we bound the gap between the performance of our IF architecture and the outage capacity, drawing upon recent IF source~\cite{de17} and channel coding~\cite{de18} bounds of the same nature. Overall, this bound and our simulations show that IF is competitive with respect to WZ in terms of end-to-end rates on uplink C-RANs. This observation, combined with the fact that IF is parallelizable (save for a final matrix multiplication), makes it an appealing alternative to WZ C-RAN architectures.

\subsection{Paper Organization}
The rest of the paper is organized as follows. Section \ref{sec: problem statement} gives a problem statement. Section \ref{sec: conventional compression} reviews
conventional techniques for distributed compression and Section \ref{sec: lattice compression} describes techniques for IF source coding. Next, Section \ref{sec: channel coding} reviews both conventional and IF channel coding, Section \ref{sec: IF CRAN architecture} presents our IF architecture for uplink C-RANs, Section \ref{sec: IF outage bound} provides a constant gap result for the outage of the proposed IF architecture and Section \ref{sec: IF CRAN algorithms} gives algorithms to optimize the associated parameters. Finally, Section \ref{sec: simulation results} presents our simulation results and Section \ref{sec: conclusion} concludes the paper.

\subsection{Notation} 
We denote column vectors by boldface lowercase (e.g., $\mathbf{x}$) and matrices by boldface uppercase (e.g., $\mathbf{X}$). Let $\|\mathbf{x}\|$ denote the Euclidean norm of the vector $\mathbf{x}$. For a matrix $\mathbf{X}$, let $\mathbf{X}^{{\T}}$ denote its transpose, $|\mathbf{X}|$ its determinant, $\mathbf{X}^{-1}$ its inverse, and $\mathbf{X} \odot \mathbf{X}$ as the elementwise square. Furthermore, let $\mathbf{X}_{\mathcal{A},\mathcal{B}}$ be the submatrix of $\mathbf{X}$ composed of the rows and columns whose indices fall the sets $\mathcal{A}$ and $\mathcal{B}$, respectively. If $\mathcal{A}=\mathcal{B}$, we write $\mathbf{X}_{\mathcal{A},\mathcal{B}}$ as $\mathbf{X}_\mathcal{A}$. All logarithms are taken to base $2$ and we define $\log^+(x)\triangleq \max (0,\log(x))$. 

\section{Problem Statement}\label{sec: problem statement}

\subsection{System Model}

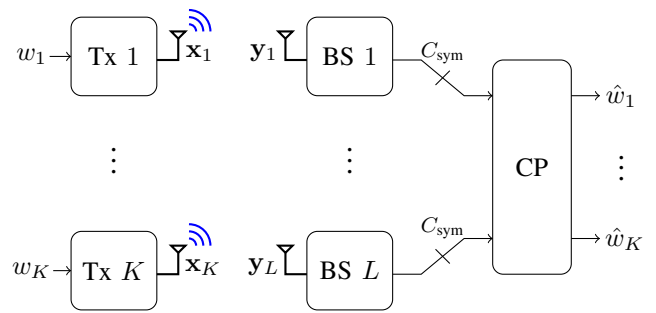
\begin{figure}[h]
\centering
{\begin{tikzpicture}[scale=0.95]

\draw [->] (-3.6,1.05)--(-3.3,1.05);
\draw [->] (-3.55,-1.95)--(-3.3,-1.95);

\node at (-3.85,1.05) {{${w}_1$}};
\node at (-3.85,-1.95) {{${w}_K$}};

\draw [black,rounded corners] (-3.3,0.5) rectangle (-2.1,1.6);
\node at (-2.7,1.05) {{Tx $1$}};
\draw [black,rounded corners] (-3.3,-2.5) rectangle (-2.1,-1.4);
\node at (-2.7,-0.3) {\large\vdots};
\node at (-2.7,-1.95) {Tx $K$};
\node at (-1.45,-1.9){{$\mathbf{x}_K$}};
\node at (-1.5,1.1){{$\mathbf{x}_1$}};

\node at (-0.6,-1.9){{$\mathbf{y}_L$}};
\node at (-0.6,1.1){{$\mathbf{y}_1$}};

\foreach \x in {-2.1}
   \foreach \y in {1,-2}
      { \draw [thick] (\x,\y)--(\x+0.3,\y)--(\x+0.3,\y+0.3);
        \draw [black, thick] (\x+0.3,\y+0.3)--(\x+0.2,\y+0.4)--(\x+0.4,\y+0.4)--(\x+0.3,\y+0.3);

	\draw[ blue, thick] (\x+0.45,\y+0.5) to [out=0,in=90] (\x+0.55, \y+0.4);
      \draw[ blue, thick] (\x+0.45,\y+0.6) to [out=0,in=90] (\x+0.65, \y+0.4);
     \draw[ blue, thick] (\x+0.45,\y+0.7) to [out=0,in=90] (\x+0.75, \y+0.4);
}

\foreach \x in {0}
   \foreach \y in {1,-2}
      { \draw [ thick] (\x,\y)--(\x-0.3,\y)--(\x-0.3,\y+0.3);
        \draw [black, thick] (\x-0.3,\y+0.3)--(\x-0.4,\y+0.4)--(\x-0.2,\y+0.4)--(\x-0.3,\y+0.3);
}

\draw [black,rounded corners] (0,0.5) rectangle (1.2,1.6);
\draw [black,rounded corners] (0,-2.5) rectangle (1.2,-1.4);

\node at (0.6,1.05) {{BS $1$}};
\node at (0.6,-0.3) {\large\vdots};
\node at (0.6,-1.95) {{BS $L$}};

\draw [->] (1.2,1) -- (1.6,1)--(2.2,0.5)--(2.6,0.5);
\draw [->] (1.2,-2) -- (1.6,-2)--(2.2,-1.5)--(2.6,-1.5);


\draw [-] (1.8,0.65)--(2.0,0.85);
\draw [-] (1.8,-1.65)--(2.0,-1.85);
\node at (1.9,1.1) {\footnotesize{$C_\text{sym}$}};
\node at (1.9,-1.35) {\footnotesize{$C_\text{sym}$}};

\draw [black,rounded corners] (2.6,-2) rectangle (3.7,1);
\node at (3.15,-0.5) {CP};

\draw [->] (3.7,0.5)--(4.1,0.5);
\draw [->] (3.7,-1.5)--(4.1,-1.5);

\node at (4.4,-0.4) {\large$\vdots$};
\node at (4.4,0.5) {{$\hat{w}_1$}};
\node at (4.45,-1.5) {{$\hat{w}_K$}};
\end{tikzpicture}}
\vspace{0.1in}
\caption{C-RAN architecture with $K$ users and $L$ BSs.}
\label{system_model}
\end{figure}

Consider the uplink C-RAN scenario illustrated in Figure~\ref{system_model}, where a set $\mathcal{K} \triangleq \{1,\ldots,K\}$ of single-antenna users communicate to a set $\mathcal{L} \triangleq \{1,\ldots,L\}$ of single-antenna base stations\footnote{For simplicity, we assume single-antenna BSs, however, the proposed schemes can be extended directly to deal with multiple-antenna BSs \cite{bn17}.}. Each BS is connected to the CP via a noiseless fronthaul links with capacity $C_\text{sym}$. The $\kth$ user encodes its message $w_k \in \{1,2,\ldots,2^{TR}\}$, with symmetric rate $R$, into a length-$T$ codeword $\mathbf{x}_k \triangleq [x_k(1)~\cdots~x_k(T)]^\T \in \mathbb{R}^{T}$ satisfying the standard power constraint $\| \mathbf{x}_k\|^2 \leq T P$. The $\lth$ BS receives $y_\ell(t) \in \mathbb{R}$ at time $t$ and the vector of all received signals $\mathbf{y}(t) \triangleq [y_1(t)~\cdots~y_L(t)]^\T $ at time $t$ can be expressed as \begin{align}\label{channel}
\mathbf{y}(t) = \mathbf{H} \mathbf{x}(t) + \mathbf{z}(t) 
\end{align} where $\mathbf{H} \in \mathbb{R}^{L \times K}$ is the channel matrix, $\mathbf{x}(t) = [x_1(t) ~ \cdots ~ x_K(t)]^\T$ is the vector of transmitted symbols at time $t$, and $\mathbf{z}(t)$ is i.i.d.~$\mathcal{N}(0,1)$. For simplicity, we focus on real-valued channels and note that complex-valued channels can be handled via their real-valued decompositions~\cite{zneg14}. 

We focus on the slow fading setting where the channel matrix is generated randomly and held fixed across all $T$ symbols. We assume that no CSI is available at the transmitters and that CSI at the receivers (CSIR) may be either global or local.  In the global CSIR scenario, the channel matrix $\mathbf{H}$ is known to the CP and all the BSs. In the local CSIR scenario, $\mathbf{H}$ is only fully known to the CP and the $\lth$ BS only knows the $\lth$ row of $\mathbf{H}$ (i.e. the channel from all users to itself). This scenario may be encountered in delay-sensitive applications where sending $\mathbf{H}$ back to the BSs through the fronthaul links is not feasible. 

The $\lth$ BS maps its observation $\mathbf{y}_\ell \triangleq [y_\ell(1)~\cdots~y_\ell(T)]^\T$ to an index $v_\ell \in  \{1,\ldots,2^{T\C}\}$ and forwards it to the CP through the fronthaul link. Upon receiving indices $v_1,\ldots,v_L$, the CP uses these indices to make estimates $\hat{w}_1,\ldots, \hat{w}_K$ of the transmitted messages. 

We say that a symmetric rate $R$ is achievable if, for any $\epsilon > 0$ and $T$ large enough, there exists encoders and decoders that can attain average probability of error at most $\epsilon$. Since we assume that $\mathbf{H}$ is only known to the receivers, each user has to tolerate some outage probability. 

\definition For a target symmetric rate $R$, we define the \textit{outage probability} of a scheme as $p_{\text{scheme}}(R) \triangleq \mathbb{P}\big(R_{\text{scheme}}(\mathbf{H}) < R\big)$ where $R_{\text{scheme}}(\mathbf{H})$ is the achievable symmetric rate under $\mathbf{H}$ for this particular scheme. Similarly, for a target outage probability $\rho$, we define the \textit{symmetric outage rate} as $R_{\text{scheme,out}}(\rho) \triangleq \sup \big\{ R : p_{\text{scheme}}(R) \leq \rho \big\}$.

\subsection{Compression-Based Strategies}
In compression-based strategies, each BS uses the fronthaul link to send a compressed version of its observation to the CP rather than decoding locally, and can thus be oblivious to the codebooks employed by the users. 
Specifically, the $\lth$ BS maps its received signal $\mathbf{y}_\ell \triangleq [y_\ell(1)~\cdots~y_\ell(T)]^\T$ to an index $v_\ell \in \{1,\ldots,2^{TR_\ell^s}\}$, where $R_\ell^s$ is the compression rate, and forwards it to the CP through a fronthaul link with capacity $\C$. Upon receiving indices $v_1,\ldots,v_L$, the CP first reconstructs the signals $\mathbf{\widehat{y}}_1,\ldots,\mathbf{\widehat{y}}_L$ where  $\widehat{\mathbf{y}}_\ell \triangleq [\hat{y}_\ell(1) ~\cdots~ \hat{y}_\ell(T)]$, then uses these reconstructions to make estimates $\hat{w}_1,\ldots, \hat{w}_K$ of the transmitted messages. Due to the limited fronthaul capacity, each decompressed signal $\hat{y}_\ell(t)= y_\ell(t) + q_\ell(t),~  \forall \ell \in \mathcal{L}$ suffers from a quantization noise $q_\ell(t)$, which is characterized via its mean-squared error (MSE) (i.e., distortion level)
$ \frac{1}{T} \mathbb{E} \big[\sum\limits_{t=1}^T (q_\ell(t))^2$\big],
which depends on the fronthaul link capacity $\C$ and the compression strategy. We assume that the $\hat{y}_\ell(t)$ are unbiased estimates of $y_\ell(t)$, since this facilitates the interface between source and channel coding by allowing the latter to assume that the quantization noise is uncorrelated with the transmitted codewords.

The end-to-end effective channel can be written as 
\begin{align}
\widehat{\mathbf{y}}(t)=\mathbf{H} \mathbf{x}(t) + \mathbf{z}(t) + \mathbf{q}(t),
\end{align}
where $\widehat{\mathbf{y}}(t) \triangleq [\widehat{y}_1(t)~\cdots~ \widehat{y}_L(t)]^\T$ and $\mathbf{q}(t) \triangleq [q_1(t) ~\cdots~ q_L(t)]^\T$. 

\remark
 Recall that the end-to-end symmetric transmission rate is denoted by $R_{\text{scheme}}(\mathbf{H})$. To avoid confusion, we write the compression rates required to convey the BSs observations to the CP with distortion levels $d_1,\ldots,d_L$ via a particular scheme by $R_{\text{scheme},\ell}^s(\mathbf{H},d_1,\ldots,d_L)$.

We say that the compression rates $R^s_{\text{scheme},\ell}(\mathbf{H},d_1,\ldots,d_L) \leq \C$ for $\ell \in \mathcal{L}$ are achievable for distortions $d_1,\ldots,d_L$, if for any $\epsilon > 0$ and $T$ large enough, there exist encoders and a decoder such that $\frac{1}{T} \mathbb{E} \big[ \sum_{t=1}^T \big(q_\ell(t)\big)^2 \big] \leq d_\ell + \epsilon$ for $\ell \in \mathcal{L}$. We also consider schemes with symmetric distortion levels $d_1=\cdots=d_L=d$ and symmetric rates $R^s_{\text{scheme},1}(\mathbf{H},d)=\cdots=R^s_{\text{scheme},L}(\mathbf{H},d)=R^s_\text{scheme}(\mathbf{H},d)$.
 
\subsection{Local CSIR vs Global CSIR} 
For global CSIR, the BSs can set the distortion levels $d_1,\ldots,d_L$ such that the required quantization rates match the fronthaul capacity, $R^s_{\text{scheme},\ell}(\mathbf{H},d_1,\ldots,d_L)=\C, \ \forall \ell \in \mathcal{L}.$ However, for the case of local CSIR, this approach is not viable since the BSs do not know the correlations between their observations. We suggest two possible workarounds. In the first, we apply ``single-user'' compression so that the quantized observation from the $\lth$ BS can be reconstructed directly from the index $v_\ell$. Unfortunately, this compression scheme does not exploit the correlation between BSs observations. In the second workaround, the BSs fix the distortion levels (regardless of $\mathbf{H}$) so that the CP can recover $\widehat{y}_1(t),\ldots,\widehat{y}_L(t)$ with some positive probability. Specifically, define the {\it compression outage probability} of a scheme, for a target distortion $d_t$ as $p^s_\text{scheme} (d_t) \triangleq \mathbb{P} \left( \max_\ell R^s_{\text{scheme},\ell}(\mathbf{H},d_t ) > \C \right)$ where the probability is taken with respect to the channel matrix $\mathbf{H}$. Note that we opt to use a symmetric distortion level $d_t$ for local CSIR, since it is not clear how to set $d_1,\ldots,d_L$ to obtain a given outage probability. For a target compression outage probability $\rho_s < \rho$, where $\rho$ is the end-to-end outage probability, the distortion achievable by a compression scheme is given by $ d_\text{scheme}(\rho_s) \triangleq \inf \left\{ d : p^s_\text{scheme} (d) \leq \rho_s \right\}$.

\section{Conventional Distributed Compression}\label{sec: conventional compression}

We now review conventional schemes for compressing the BSs observations' $\mathbf{y}_1,\ldots,\mathbf{y}_L$.

\subsection{``Single-User'' Compression}\label{conventional_SUC}

The simplest compression strategy is to ignore any correlations between $\mathbf{y}_1 ,\ldots, \mathbf{y}_L$. The $\lth$ BS quantizes its observation $\mathbf{y}_\ell$ and sends its index $v_\ell$ to the CP through the fronthaul link. Then, the CP independently recovers $\widehat{\mathbf{y}}_\ell$ using $v_\ell$ for $\ell=1,\ldots,L$. Using i.i.d.~Gaussian codebooks, the following rates are achievable.

\begin{lemma}[{\cite[Equation (8)]{YW16}}]
The achievable compression rates for single-user compression (SUC) are 
\begin{align}\label{R_SU}
R^s_{\text{SU},\ell}(\mathbf{H}_{\ell,\mathcal{K}},d_\ell) = \frac{1}{2} \log \left( 1 + \frac{P \|\mathbf{H}_{\ell,\mathcal{K}}\|^2 +1   }{d_\ell}	 \right), \ \  \forall \ell \in \mathcal{L}
\end{align}
where $\mathbf{H}_{\ell,\mathcal{K}}$ denotes the $\lth$ row of the channel matrix $\mathbf{H}$ and $d_\ell$ is the $\lth$ distortion level.
\end{lemma} Note that the ``$1~+$'' appears inside the logarithm since we insist upon unbiased estimates.

For symmetric fronthaul constraints $R^s_{\text{SU},\ell}(\mathbf{H}_{\ell,\mathcal{K}},d_{\text{SU},\ell}) \leq C_\text{sym}$, the optimal distortions $d_{\text{SU},\ell}$ under SUC are 
\begin{align}\label{SUC_dist_levels}
d_{\text{SU},\ell}=\frac{\|\mathbf{H}_{\ell,\mathcal{K}}\|^2 P+1}{2^{2C_\text{sym}}-1}, \ \ \forall \ell \in \mathcal{L}.
\end{align}

\subsection{Wyner-Ziv Compression}

Since the received signals $\mathbf{y}_1, \ldots, \mathbf{y}_L$ are correlated, we can use a Wyner-Ziv (WZ) compression strategy~\cite{wz76} to exploit this correlation using successive decompression~\cite{zy13,psss13,psss14,zy14,YW16,ZXYC16}. Assume the decompression order is specified by a permutation $\pi_s: \mathcal{L} \rightarrow \mathcal{L}$. The basic idea behind WZ strategy is for the CP to use previously decompressed signals $\widehat{\mathbf{y}}_{\pi_s(1)},\ldots, \widehat{\mathbf{y}}_{\pi_s(\ell-1)}$ as side information while recovering $\widehat{\mathbf{y}}_{\pi_s(\ell)}$ to obtain a finer reconstruction.

\begin{lemma}[{\cite[Equation (28)]{YW16}}]\label{WZ_lemma}
The achievable compression rates under Wyner-Ziv compression are given by
\begin{align}\label{WZ_rates_global}
&R^s_{\text{WZ},\pi_s(\ell)} (\mathbf{H},\mathbf{D}_{\mathcal{T}_\ell}) =\\  &\scalemath{0.98}{\frac{1}{2} \log \left( \frac{ \big| P \mathbf{H}_{\mathcal{T}_\ell,\mathcal{K}} (\mathbf{H}_{\mathcal{T}_\ell,\mathcal{K}})^\T  + \mathbf{I} + \mathbf{D}_{\mathcal{T}_\ell}   \big| }{\big|  P \mathbf{H}_{\mathcal{T}_{\ell-1},\mathcal{K}} (\mathbf{H}_{\mathcal{T}_{\ell-1},\mathcal{K}})^\T  + \mathbf{I} + \mathbf{D}_{\mathcal{T}_{\ell-1}}  \big| }	 \right) - \frac{1}{2} \log ( d_{\pi_s(\ell)} )} \nonumber
\end{align} for $\ell \in \mathcal{L}$ where $\mathcal{T}_\ell  \triangleq  \{\pi_s(1),\ldots,\pi_s(\ell)\}$ and $\mathbf{D} \triangleq \text{diag}(d_1,\ldots,d_L)$.
\end{lemma} See \cite{YW16} for a proof.

{\remark[Global CSIR]
It can be shown that, for a fixed $\mathbf{D}_{\mathcal{T}_\ell-1}$, $R^s_{\text{WZ},\pi_s(\ell)} (\mathbf{H},\mathbf{D}_{\mathcal{T}_\ell}) $ is monotonically decreasing in $d_{\pi_s(\ell)}$. This means that the optimal $d_{\text{WZ},\pi_s(\ell)}$ for the global CSIR case can be obtained successively for $\ell=1,\ldots,L$ (e.g., using a bisection search method) such that $R^s_{\text{WZ},\pi_s(\ell)} (\mathbf{H},\mathbf{D}_{\mathcal{T}_\ell}) = C_\text{sym}$.}

{\remark[Local CSIR]
Under local CSIR, $\mathbf{H}$ is only known to the CP. Thus, the rates in \eqref{WZ_rates_global} are not known to the BSs and we cannot set the distortion levels accordingly. Furthermore, it is not clear how to optimize for the asymmetric distortion levels to satisfy a certain outage probability $\rho_s$. Hence, we set a symmetric distortion $d_t$ such that $ \mathbb{P}\left( \max_\ell R^s_{\text{WZ},\ell}(\mathbf{H},d_t\mathbf{I}) > \C \right) \leq \rho_s. $}

\subsection{Symmetric Berger-Tung Compression}
The rate region for distributed Gaussian source coding remains an open problem. However, the Berger-Tung (BT) quantize-and-bin strategy~\cite{tungphd} is known to be optimal for two (scalar) sources~\cite{wtv08}. Here, following the example of~\cite{oe17}, we take the BT rate region, evaluated for Gaussian test channels and with a symmetric distortion constraint, as a benchmark for our compression schemes. This strategy relies upon simultaneous joint typicality decoding, which has substantially higher implementation complexity than the sequential decoding used for WZ compression.

\begin{lemma}\label{BT}
The achievable symmetric compression rate using the BT compression scheme is
\begin{align}\label{BT_symmetric_rate}
R^s_\text{BT}(\mathbf{H},d_\text{BT}) = \frac{1}{2L} \log \left| \mathbf{I} + \frac{1}{d_\text{BT}} \mathbf{K}_{YY}  \right|
\end{align} where $\mathbf{K}_{YY}= P \mathbf{H} \mathbf{H}^\T + \mathbf{I}$ is the covariance matrix of the BS observations and $d_\text{BT}$ is the symmetric distortion level. 
\end{lemma} See~\cite[Sec. II]{oe17} for further details.

{\remark[Global CSIR]	
Similar to the WZ scheme, $R^s_\text{BT}(\mathbf{H},d_\text{BT})$ is monotonically decreasing in $d_\text{BT}$, thus we can use a bisection search method to choose $d_\text{BT}$ such that $R^s_\text{BT}(\mathbf{H},d_\text{BT})=C_\text{sym}.$}

{\remark[Local CSIR] For local CSIR, one can still implement the BT quantize-and-bin strategy by fixing $d_\text{BT}$, independent of the channel $\mathbf{H}$, to the smallest value that satisfies $p^s_\text{BT}\leq \rho_s$.

\section{Lattice Distributed Compression}\label{sec: lattice compression}
We now describe schemes for lattice quantization for both global and local CSIR scenarios. We start with definitions and coding theorems for lattices that will be useful in our strategies.

\subsection{Lattice Preliminaries} \label{sec: lattice preliminaries}
A \textit{lattice} $\Lambda$ is a discrete additive subgroup of $\mathbb{R}^T$ that is closed under addition and reflection, and can be expressed as $\Lambda = \{  \mathbf{F} \mathbf{w} : \mathbf{w} \in \mathbb{Z}^T \}$ for some (non-unique) generator matrix $\mathbf{F} \in \mathbb{R}^{T \times T}$. The \textit{lattice quantizer} maps any point in $\mathbb{R}^T$ to the nearest point in $ {\Lambda}$ (breaking ties systematically),  
$$\mathcal{Q}_{\Lambda} (\mathbf{x}) \triangleq \argmin\limits_{\boldsymbol{\lambda} \in \Lambda} \|\mathbf{x}-\boldsymbol{\lambda}\|^2,$$ which in turn defines the \textit{Voronoi region} $\mathcal{V}(\Lambda)$  as the subset of $\mathbb{R}^T$ that quantizes to the zero vector. The $\bmod \ \Lambda $ operator returns the lattice quantization error $[\mathbf{x}] \bmod{\Lambda} \triangleq \mathbf{x}-\mathcal{Q}_{\Lambda}(\mathbf{x})$ and the \textit{second moment} of a lattice is $ \sigma^2 (\Lambda) \triangleq \frac{1}{T} \mathbb{E}\| \mathbf{x} \|^2 $ for $\mathbf{x} \sim \text{Unif} \left( \mathcal{V}\left( \Lambda \right) \right)$. We also define the \textit{dual lattice} as $\Lambda^* \triangleq \{ \mathbf{F}^{-\T} \mathbf{a}: \mathbf{a} \in \mathbb{Z}^{K} \}$.

Let $\mathcal{B}(\mathbf{0},r) \triangleq \{\mathbf{x} \in \mathbb{R}^K : \|\mathbf{x}\| \leq r\}$ be the ball centered at the origin with radius $r$. For a lattice $\Lambda =  \{ \mathbf{F} \mathbf{a}: \mathbf{a} \in \mathbb{Z}^{K} \}$, the \textit{$\mth$ successive minima} is  \begin{align}{\lambda}_m(\mathbf{F}) \triangleq \inf\{ r > 0: \Lambda\, \cap \, &\mathcal{B}(\mathbf{0},r) \text{~contains $m$ linearly} \\ &~~\text{independent lattice points}\}\nonumber \end{align} for $m=1,\ldots,K$. The following transference theorem due to Banaszczyk allows us to connect the successive minima of a lattice to those of its dual. 
\lemma[{{\cite[Theorem 2.1]{Ba93}}}] \label{thm:transference}For $m = 1,\ldots,K$, we have that $\lambda_m(\mathbf{F})\lambda_{K-m+1}(\mathbf{F}^{-\T}) \leq K$.

\lemma[Crypto Lemma]\label{crypto_lemma}  For $\mathbf{y} \in \mathbb{R}^T$ and a dither $\mathbf{u}\sim \text{Unif}(\mathcal{V}(\Lambda))$ independent of $\mathbf{y}$, we have that $\mathbf{q} = \left[ \mathbf{y+u} \right] \bmod \Lambda$ is independent of $\mathbf{y}$ and $\mathbf{q} \sim \text{Unif}(\mathcal{V}(\Lambda))$. See \cite{zamir} for a proof.

We say that the lattice $\Lambda_{C}$ is nested in the lattice $\Lambda_F$ if $\Lambda_C \subset \Lambda_F$. A nested lattice codebook $\Lambda_F \cap \mathcal{V}(\Lambda_C)$ consists of all the fine lattice points inside the fundamental Voronoi region of the coarse lattice.  Note that nested lattices $\Lambda_C \subset \Lambda_F$ satisfy a distributive law, i.e., for any $\mathbf{x},\mathbf{y} \in \mathbb{R}^T$ and $a,b \in \mathbb{Z}$, 
\begin{align} &\left[ a \left[ \mathbf{x} \right] \bmod \Lambda_C + b \left[ \mathbf{y} \right] \bmod \Lambda_C \right] \bmod \Lambda_F \label{distributivelaw} \\ &=\left[ a \mathbf{x}+b \mathbf{y} \right] \bmod \Lambda_F. \nonumber \end{align}

The following lemma encapsulates some of the nested lattice existence results from~\cite{oe16} in a form suitable for establishing our integer-forcing source coding results.

\lemma[{\cite[Theorem 2]{oe16}}] \label{good_nested_lattices}  
For $\theta_1,\ldots,\theta_{K} \in \mathbb{R}, \epsilon>0$, and $T$ large enough, there exist a nested lattice chain $\Lambda_{K}  \subseteq \ldots\subseteq \Lambda_{1}$ (generated using Construction A from a $p$-$\text{ary}$ linear code for a large enough prime $p$) such that for $m = 1,\ldots,L$,
\begin{itemize}
\item [1.] The second moment satistfies $\theta_m \leq \sigma^2(\Lambda_{m}) < \theta_m + \epsilon$.
\item [2.] A mixture of Gaussian and lattice quantization noise will remain in the Voronoi cell w.h.p. if its second moment is below $\theta_m$. Specifically,  if $\mathbf{z}_\text{eff}= \beta_0 \mathbf{z}_0 + \sum\limits_{k=1}^{K} \beta_n \mathbf{z}_k$ where $\beta_0,\ldots,\beta_{K} \in \mathbb{R}$, $\mathbf{z}_0 \sim \mathcal{N} ( \mathbf{0},\mathbf{I})$, $\mathbf{z}_k \sim \text{Unif}(\mathcal{V}(\Lambda_{k}))$ and if $\beta_0^2+\sum\limits_{k=1}^{K} \beta_k^2 \theta_k < \theta_m$, then $\Pr \left( [\mathbf{z}_\text{eff}] \bmod \Lambda_m \neq \mathbf{z}_\text{eff} \right) \leq \epsilon$.
\item[3.]  The rate of the codebook formed by intersecting $\Lambda_m$ with $\mathcal{V}(\Lambda_\ell)$ satisfies $$\frac{1}{2} \log \left( \frac{\theta_\ell}{\theta_m} \right) \leq \frac{1}{T} \log \big| \Lambda_m  \cap \mathcal{V}(\Lambda_\ell) \big| < \frac{1}{2} \log \left( \frac{\theta_\ell}{\theta_m} \right) +\epsilon. $$ 
\end{itemize}

\subsection{Integer-Forcing Source Coding with Global CSIR}\label{sec: sym IFSC}

The integer-forcing source coding (IFSC) strategy exploits
the correlation between $\mathbf{y}_1,\ldots,\mathbf{y}_L$ by first recovering integer-linear combinations
\begin{align}
\mathbf{v}_{s,m} \triangleq \sum\limits_{\ell=1}^{L} {a}_{s,m,\ell}  \, \widehat{\mathbf{y}}_\ell   , \ \ \ {a}_{s,m,\ell} \in \mathbb{Z}, \ \forall m \in \mathcal{L}.
\end{align}

As we will see, by optimizing over the choices of these integers, we can reduce the variances of the linear combinations $\mathbf{v}_{s,1},\ldots,\mathbf{v}_{s,L}$, thus relaxing the requirements on the second moment of the coarse lattice and decreasing the compression rate. If we recover $L$ linearly independent integer combinations, we can solve them for the quantized sources $\widehat{\mathbf{y}}_1,\ldots,\widehat{\mathbf{y}}_L$.

Let $\mathbf{A}_s$ be the $L \times L$ integer matrix whose $(m,\ell)^\text{th}$ entry is $a_{s,m,\ell}$ and note that we can solve for the quantized sources if $\mathbf{A}_s$ is full rank.

\subsubsection*{\bf{Codebook}} We use a nested lattice codebook $\mathcal{C}\triangleq \Lambda_F \cap \mathcal{V}(\Lambda_C)$ comprised of nested lattices $\Lambda_C \subset \Lambda_F$ selected using Lemma \ref{good_nested_lattices} with rate $\C$ and parameters $\theta_F = d$ and $\theta_C = d \, 2^{2 \C}$ where $d$ is the achievable symmetric distortion to be set later.

\subsubsection*{\bf{Compression}} The $\lth$ BS adds a random dither $\mathbf{u}_\ell \sim \text{Unif}(\mathcal{V}(\Lambda_{F}))$, then computes
\begin{align}
\boldsymbol{\lambda}_\ell = \left[ \mathcal{Q}_{\Lambda_F} \left( \mathbf{y}_\ell + \mathbf{u}_\ell\right) \right] \bmod \Lambda_C.
\end{align}Note that the random dithers $\mathbf{u}_1,\ldots,\mathbf{u}_L$ are independent and known to the CP\footnote{The availability of random dithers at the transmitters and receivers is a standard assumption made to streamline achievability proofs for nested lattice codes. It is straightforward to show that the same rates are achievable by replacing these random dithers with deterministic ones. See, for instance,~\cite[App. C]{ng11IT} for more details.}. The $\lth$ BS then sends the index $v_\ell \in \left\{ 1,\ldots,2^{T\C} \right\}$ of the codeword $\boldsymbol{\lambda}_\ell$ to the CP through the fronthaul link.

\subsubsection*{\bf{Decompression}}
 For each $\ell \in \mathcal{L}$, the CP first recovers $\boldsymbol{\lambda}_\ell$ from $v_\ell$, then removes the dithers to recover  \begin{align*}
\widetilde{\mathbf{y}}_\ell &=\left[ \boldsymbol{\lambda}_\ell - \mathbf{u}_\ell \right] \bmod \Lambda_{C}\\ &\stackrel{(a)}{=} \left[  \mathcal{Q}_{\Lambda_{F}} \left(  {\mathbf{y}}_\ell +\mathbf{u}_\ell \right)    -\mathbf{u}_\ell \right] \bmod \Lambda_{C}\\ &\stackrel{(b)}{=}\left[  {\mathbf{y}}_\ell + {\mathbf{q}}_\ell \right] \bmod \Lambda_{C}
\end{align*}
where $(a)$ holds due to the distributive law, $(b)$ holds from ${\mathbf{q}}_\ell = -[ \mathbf{y}_\ell+\mathbf{u}_\ell]\bmod \Lambda_F$ which is independent of ${\mathbf{y}}_\ell$ and uniformly distributed over $\mathcal{V}(\Lambda_F)$ by the Crypto Lemma.
 
 The CP then estimates the integer-linear combinations
\begin{align}\label{s_estimate}
\widehat{\mathbf{v}}_{s,m} &=  \left[ \sum\limits_{\ell=1}^{L} {a}_{s,m,\ell} \widetilde{\mathbf{y}}_\ell \right] \bmod{\Lambda_C} \\& \stackrel{(a)}{=}  \left[ \sum\limits_{\ell=1}^{L} {a}_{s,m,\ell}  \left( {\mathbf{y}}_\ell + {\mathbf{q}}_\ell \right)  \right] \bmod{\Lambda_C} \\&\stackrel{\text{w.h.p.}}{=} \sum\limits_{\ell=1}^{L} {a}_{s,m,\ell} \left( {\mathbf{y}}_\ell + \mathbf{q}_\ell \right) 
\end{align}
where $(a)$ holds from the distributive law and the last equality holds with high probability by the second property of Lemma \ref{good_nested_lattices} if the effective variance satisfies $ \frac{1}{T} \mathbb{E} \big[ \|  \sum\limits_{\ell=1}^{L} {a}_{s,m,\ell} \left( {\mathbf{y}}_\ell +\mathbf{q}_\ell \right)  \|^2 \big] < \theta_C$ for $\ell=1,\ldots,L$. This can be guaranteed by choosing $d$ such that $$ \max_\ell \mathbf{a}_{s,\ell}^\T \left( \mathbf{K}_{YY}+d\mathbf{I} \right) \mathbf{a}_{s,\ell} +\epsilon =\theta_C= d~2^{2\C}$$ where $\mathbf{K}_{YY} \triangleq \frac{1}{T} \mathbb{E}\left[ \mathbf{Y} \mathbf{Y}^\T \right] = P\mathbf{HH}^\T +\mathbf{I}$ is the effective covariance of $\mathbf{Y}\triangleq [\mathbf{y}_1~\cdots~\mathbf{y}_L]^\T$, $ \frac{1}{T} \mathbb{E}\|\mathbf{q}_\ell\|^2 = d$, $ \mathbf{a}_{s,m} \triangleq [ a_{s,m,1}~\cdots~a_{s,m,L}]^\T$, and $\epsilon > 0$ (and will be sent to $0$ as the block length $T$ tends to infinity).

Assuming correct recovery $\widehat{\mathbf{v}}_{s,m}=\mathbf{v}_{s,m}$ for $m=1,\ldots,L$, the CP forms the matrix $$\mathbf{V}_s \triangleq [\mathbf{v}_{s,1}~\cdots~\mathbf{v}_{s,L}]^\T= \mathbf{A}_s \left( {\mathbf{Y}}+\mathbf{Q} \right)$$ where ${\mathbf{Q}} \triangleq [{\mathbf{q}}_1~\cdots ~{\mathbf{q}}_{L}]^\T$, then applies the inverse of $\mathbf{A}_s$ to obtain $ \widehat{\mathbf{Y}} \triangleq \mathbf{A}_s^{-1} \mathbf{V}_s = {\mathbf{Y}}+ \mathbf{Q}.$

\begin{lemma}[{\cite[Theorem 1]{oe17}}]  The IFSC symmetric rate for symmetric distortion $d$ is 
\begin{align}\label{IFSC_rate1}
R^s_\text{IFSC} (\mathbf{H},d) = \min\limits_{\mathbf{A}_s \in \mathbb{Z}^{L \times L}}   \max\limits_{ \ell=1,\ldots,L}  \frac{1}{2} \log^+ \| \mathbf{F}_s \mathbf{a}_{s,\ell} \|^2
\end{align} where $\mathbf{F}_s$ is any matrix satisfying $\mathbf{F}_s^\T \mathbf{F}_s =\frac{1}{d} \mathbf{K}_{{Y}{Y}} +  \mathbf{I}$, $\mathbf{K}_{YY}=P\mathbf{H}\mathbf{H}^\T +\mathbf{I}$ is the effective covariance matrix of ${\mathbf{Y}}$ and the minimization is over all full-rank $\mathbf{A}_s \in \mathbb{Z}^{L \times L}$.
\end{lemma}

\remark It may seem that the lattice codebook $\mathcal{C}$ depends on the channel realization $\mathbf{H}$, since $\theta_F=d$ and $d$ should be set such that $R_\text{IFSC}^s(\mathbf{H},d)=\C$. However, we can fix a codebook $\mathcal{C}$ with arbitrary $d'>0$, independent of $\mathbf{H}$, and scale it (i.e., scale both $\Lambda_F$ and $\Lambda_C$) using a parameter $\beta$ that depends on $\mathbf{H}$, such that $d=\beta^2 d'$. It is also worth noting that with asymmetric scaling parameters $\beta_1,\ldots,\beta_L$ and setting $\mathbf{A}_s=\mathbf{I}$, we can recover the SUC performance of the i.i.d.~Gaussian codebooks from~\eqref{SUC_dist_levels}.

\subsection{Asymmetric Integer-Forcing Source Coding with Global CSIR}
We now recall the asymmetric IFSC strategy introduced in \cite{HN16}. 

\subsubsection*{\bf{Codebook}}
Without loss of generality, assume we have $2L$ nested lattices $\Lambda_{C,L} \subseteq \cdots \subseteq \Lambda_{C,1} \subseteq \Lambda_{F,L} \subseteq \cdots \subseteq \Lambda_{F,1}$ generated using Lemma \ref{good_nested_lattices} with parameters $\theta_{C,1} \leq \cdots \leq \theta_{C,L} \leq \theta_{F,1} \leq \cdots \leq \theta_{F,L}$ that each pair forms a codebook $\mathcal{C}_\ell \triangleq \Lambda_{F,\ell} \cap \mathcal{V}(\Lambda_{C,\ell})$ with rate $\frac{1}{2} \log \left( \frac{\theta_{C,\ell}}{\theta_{F,\ell}} \right) = \C$.

\subsubsection*{\bf{Compression}}
The $\lth$ BS maps $\mathbf{y}_\ell$ to a lattice codeword $\boldsymbol{\lambda}_\ell \in \mathcal{C}_\ell$ 
\begin{align}
\boldsymbol{\lambda}_\ell = \left[ \mathcal{Q}_{\Lambda_{F,\ell}} \left( \mathbf{y}_\ell +\mathbf{u}_\ell \right) \right] \bmod \Lambda_{C,\ell}, \ \ \forall \ell \in \mathcal{L}
\end{align}
where $\mathbf{u}_\ell$ is a random dither uniformly distributed over $\mathcal{V}(\Lambda_{F,\ell})$ and independent of $\mathbf{y}_\ell$.

\subsubsection*{\bf{Algebraic Successive Decompression}}
For a full-rank integer matrix $\mathbf{A}_s$, assume that the recovered combinations $\mathbf{v}_{s,1},\ldots,\mathbf{v}_{s,L}$ have been re-indexed (i.e., the rows of $\mathbf{A}_s$) such that their effective variances are monotonically increasing (i.e., $\mathbb{E}\|\mathbf{v}_{s,1}\|^2 \leq \cdots \leq \mathbb{E}\|\mathbf{v}_{s,L}\|^2$). Furthermore, assume that the BSs are re-indexed (i.e., the columns of $\mathbf{A}_s$, columns and rows of $\mathbf{K}_{YY}$ as well as the diagonal elements of $\mathbf{D}$) such that the full-rank integer matrix $\mathbf{A}_s$ has full-rank sub-matrices $\mathbf{A}_{s,[1:m]}$, for $m=1,\ldots,L$. Finally, note that for the symmetric rate constraint $\C$, the monotonically increasing effective variances $\mathbb{E}\|\mathbf{v}_{s,1}\|^2 \leq \cdots \leq \mathbb{E}\|\mathbf{v}_{s,L}\|^2$ induces the monotonically increasing distortion levels $d_1 \leq \cdots \leq d_L$.

Given the lattice codewords $\boldsymbol{\lambda}_1,\ldots,\boldsymbol{\lambda}_L$, the CP removes the dithers to get 
\begin{align}
\widetilde{\mathbf{y}}_\ell &= \left[ \boldsymbol{\lambda}_\ell - \mathbf{u}_\ell \right] \bmod \Lambda_{C,\ell}  \stackrel{(a)}{=} \left[ \mathbf{y}_\ell + \mathbf{q}_\ell \right] \bmod \Lambda_{C,\ell}
\end{align} where $(a)$ follows from the distributive law, and $\mathbf{q}_\ell=-[\mathbf{y}_\ell+\mathbf{u}_\ell] \bmod \Lambda_{F,\ell}$ is uniformly distributed over $\mathcal{V}(\Lambda_{F,\ell})$ and independent of $\mathbf{y}_\ell$ by the Crypto lemma.

\sloppy For $m=1,\ldots,L$, we attempt to recover the $\mth$ integer linear combination $\mathbf{v}_{s,m} = \sum_{\ell=1}^L a_{s,m,\ell} \left( \mathbf{y}_\ell+\mathbf{q}_\ell \right)$ using $\mathbf{v}_{s,1},\ldots,\mathbf{v}_{s,m-1}$ as side information. The main technical obstacle is that the CP has $\widetilde{\mathbf{y}}_\ell = \left[ \mathbf{y}_\ell + \mathbf{q}_\ell \right] \bmod \Lambda_{C,\ell}$, whereas we need $\widetilde{\mathbf{y}}_\ell = \left[ \mathbf{y}_\ell + \mathbf{q}_\ell \right] \bmod \Lambda_{C,m}$ for $\ell\in\mathcal{L}$ to form $ \left[ \sum_{\ell=1}^L a_{s,m,\ell}  \left( \mathbf{y}_\ell + \mathbf{q}_\ell \right) \right] \bmod \Lambda_{C,m}$. If $\Lambda_{C,\ell} \subseteq \Lambda_{C,m}$, then $ \left[ \left[ \mathbf{y}_\ell + \mathbf{q}_\ell \right] \bmod \Lambda_{C,\ell} \right] \bmod \Lambda_{C,m}= \left[ \mathbf{y}_\ell + \mathbf{q}_\ell  \right] \bmod \Lambda_{C,m}$. However, this relation does not hold for $\Lambda_{C,\ell} \supset \Lambda_{C,m}$. However, as shown in Lemma~\ref{IFSC_SIC} of Appendix~\ref{IFSC_SIC_sec}, we can use $\mathbf{v}_{s,1},\ldots,\mathbf{v}_{s,m-1}$ to recover 
\begin{align}
\mathbf{t}_{m,\ell} \triangleq \left[ \mathbf{y}_\ell + \mathbf{q}_\ell \right] \bmod \Lambda_{C,m}, \ \ \ \forall \ell \in \mathcal{L}.
\end{align}

The CP then estimates
\begin{align}
\widehat{\mathbf{v}}_{s,m} &= \left[\sum\limits_{\ell=1}^L a_{s,m,\ell} \mathbf{t}_{m,\ell} \right] \bmod \Lambda_{C,m} \\ &\stackrel{(a)}{=} \left[\sum\limits_{\ell=1}^L a_{s,m,\ell} \left( {\mathbf{y}}_\ell+\mathbf{q}_\ell \right)  \right] \bmod \Lambda_{C,m} \\ &\stackrel{(b)}{=} \mathbf{v}_{s,m} 
\end{align}
where $(a)$ holds from the distributive law and $(b)$ holds with high probability if $$\mathbf{a}_{s,m}^\T \left(\mathbf{K}_{{Y}{Y}} +  \mathbf{D} \right) \mathbf{a}_{s,m} < \theta_{C,m}$$ where $\mathbf{K}_{YY}=P\mathbf{HH}^\T+\mathbf{I}$ and $\mathbf{D}\triangleq \text{diag}({d}_{1},\ldots,{d}_{L})$. 

Finally, by setting $\theta_{F,m}=d_m$ and $\theta_{C,m}= \mathbf{a}_{s,m}^\T \left(\mathbf{K}_{YY}+\mathbf{D} \right) \mathbf{a}_{s,m} +\epsilon$ for some $\epsilon$ that tends to zero as the blocklength goes to infinity, the CP recovers $\mathbf{V}_s=\mathbf{A}_s \left(\mathbf{Y}+\mathbf{Q} \right)$ with high probability where $\mathbf{V}_s \triangleq [\mathbf{v}_{s,1}~\cdots~\mathbf{v}_{s,L}]^\T$. Finally, it applies the inverse $\mathbf{A}_s^{-1}$ to obtain $\widehat{\mathbf{Y}}\triangleq \mathbf{Y}+\mathbf{Q}$.

\begin{lemma}[{\cite[Theorem 3]{HN16}}] \label{lem:SIFSC} The achievable asymmetric rates for IFSC with algebraic successive cancellation are
\begin{align}\label{IFSC_rate_SIC}
& R^s_{\text{IFSC},\ell} (\mathbf{H},\mathbf{D}) = \min\limits_{\substack{\mathbf{A}_s \in \mathbb{Z}^{L \times L} }}   \frac{1}{2} \log^+ \left( \frac{\mathbf{a}_{s,\ell}^\T \left(\mathbf{K}_{{Y}{Y}} +  \mathbf{D} \right) \mathbf{a}_{s,\ell} }{d_\ell}  \right)
\end{align} for $\ell=1,\ldots,L$ where the minimization over all integer matrices $\mathbf{A}_s$ such that $\text{Rank}(\mathbf{A}_{s,[1:m]}) = m$ for $m=1,\ldots,L$ and 
$ \mathbf{a}_{s,1}^\T \left( \mathbf{K}_{YY}+\mathbf{D} \right) \mathbf{a}_{s,1} \leq \cdots \leq 
\mathbf{a}_{s,L}^\T \left( \mathbf{K}_{YY}+\mathbf{D} \right) \mathbf{a}_{s,L}.
$ \end{lemma} 

\remark \label{IFSC_out} Note that, to achieve the rates in \eqref{IFSC_rate1} or \eqref{IFSC_rate_SIC}, all BSs need to know $\mathbf{K}_{YY}=P\mathbf{HH}^\T+\mathbf{I}$, and thus require global CSIR. Furthermore, for asymmetric IFSC, the second moment of the fine lattices define the distortion levels directly. This means that the BSs must maintain a collection of codebooks in order to match their distortion levels $d_\ell$ to the realization of $\mathbf{H}$.

Next, we turn to discuss compression schemes suitable for local CSIR.

\subsection{Integer-Forcing Source Coding with Local CSIR} \label{sec: IFSC local csir}
In the local CSIR setting, the BSs must tolerate some probability of outage in order to exploit the correlations between $\mathbf{y}_1,\ldots,\mathbf{y}_L$. Specifically, to ensure that the CP can successfully recover $\widehat{\mathbf{y}}_1,\ldots,\widehat{\mathbf{y}}_L$ with probability approaching $1-\rho_s$, we use the IFSC scheme from Section \ref{sec: sym IFSC} with a fixed symmetric distortion $d_t$ chosen (e.g., using bisection search) such that $p^s_\text{outage}(d_t)=\rho_s$.

The compression and decompression processes are similar to the IFSC in Section \ref{sec: sym IFSC}. However, for local CSIR, we do not adapt the fine and coarse lattices according to the channel matrix $\mathbf{H}$. Rather, we select a fixed codebook to attain the desired outage probability $\rho_s$.

{ \remark
It is worth noting that the end-to-end outage event is the union of two events, namely, the event that the CP fails to recover $\widehat{\mathbf{y}}_1,\ldots,\widehat{\mathbf{y}}_L$ successfully (i.e., compression outage) and the event that the CP fails to decode the messages $w_1,\ldots,w_L$, even with a successful recovery of  $\widehat{\mathbf{y}}_1,\ldots,\widehat{\mathbf{y}}_L$ (i.e., channel coding outage). Hence, the target compression outage probability $\rho_s$ and the channel coding outage should be set such that the end-to-end outage probability does not exceed $\rho$. In our work, we simply take $\rho_s=\frac{\rho}{2}$.}

\subsection{Opportunistic IFSC for Local CSIR} \label{{sec: Opportunistic IFSC local csir}} 

For some channel realizations $\mathbf{H}$, the achievable distortion levels under SU compression $d_{\text{SU},\ell}$ in \eqref{SUC_dist_levels} may in fact be smaller than the fixed symmetric distortion $d_t$ that attains the desired outage probability $\rho_s$. This observation suggests the following opportunistic scheme that combines the IFSC and SUC schemes. First, we choose a lattice codebook with a fine lattice that induces a distortion level $d_t$ as in Section \ref{sec: IFSC local csir}. Then, for $\ell$ such that $d_{\text{SU},\ell} < d_t$, the $\lth$ BS scales its observation using a parameter $\beta_\ell$ such that the CP reconstructs $\mathbf{y}_\ell$ up to distortion $d_{\text{SU},\ell}$ before forming the linear combinations as in the IFSC scheme. For the remaining $\ell$ such that $d_{\text{SU},\ell}> d_t$, we proceed as in the basic IFSC scheme. Note that the effective variance of the combinations will be reduced. Next, we present the opportunistic scheme in detail.

\subsubsection*{\bf{Codebook}} Select a nested lattice pair $\Lambda_F \subseteq \Lambda_C$ using Lemma~\ref{good_nested_lattices} with parameters $\theta_{F} = d_t$ and $\theta_C = d_t 2^{2C_\text{sym}}$, where $d_t$ is the target symmetric distortion. The nested lattice pair forms the lattice codebook $\mathcal{C} \triangleq \Lambda_{F} \cap \mathcal{V}(\Lambda_C)$ with rate $C_\text{sym}$.

\subsubsection*{\bf{Compression}}
Using the codebook $\mathcal{C}$, $\lth$ BS maps its observation $\mathbf{y}_\ell$ to the lattice codeword
\begin{align}
\boldsymbol{\lambda}_\ell = \left[ \mathcal{Q}_{\Lambda_F} \left( \beta_\ell \mathbf{y}_\ell + \mathbf{u}_\ell\right) \right] \bmod \Lambda_C
\end{align}
where $\mathbf{u}_\ell$ is a random dither uniformly distributed over $\mathcal{V}(\Lambda_{F})$ and $\beta_\ell=1$ whenever $d_{\text{SU},\ell} > d_t$. However, when $d_{\text{SU},\ell} < d_t$ we have 
\begin{align} \label{SUC_op_1}
R^s_{\text{SU},\ell} (\mathbf{H},d_t) \triangleq \frac{1}{2} \log \left( \frac{P \|\mathbf{H}_{\ell,\mathcal{K}}\|^2+1 + d_t}{d_t} \right) < \C
\end{align} and we can better utilize the $\lth$ fronthaul link by scaling up $\mathbf{y}_\ell$ using $\beta_\ell > 1 $ such that 
\begin{align}\label{SUC_op_2}
\beta_\ell = \sqrt{\frac{d_t \left(2^{2C_\text{sym}}-1 \right)-\epsilon}{P \|\mathbf{H}_{\ell,\mathcal{K}}\|^2+1 }}.
\end{align} where $\epsilon$ goes to zero as the blocklength goes to infinity.

\subsubsection*{\bf{Decompression}}
First, the CP recovers 
\begin{align}
\widetilde{\mathbf{y}}_\ell &\triangleq \left[ \boldsymbol{\lambda}_\ell - \mathbf{u}_\ell \right] \bmod \Lambda_C \\ &\stackrel{(a)}{=}\left[ \beta_\ell \mathbf{y}_\ell +\widetilde{\mathbf{q}}_\ell \right] \bmod \Lambda_C \\ &\stackrel{(b)}{=} \begin{dcases}
		\left[  \mathbf{y}_\ell +\widetilde{\mathbf{q}}_\ell \right] \bmod \Lambda_C  & \mbox{if } d_{\text{SU},\ell} > d_t \\
		 \beta_\ell \mathbf{y}_\ell +\widetilde{\mathbf{q}}_\ell   & \mbox{if } d_{\text{SU},\ell} < d_t
	\end{dcases}
\end{align} where $\widetilde{\mathbf{q}}_\ell = -\left[ \beta_\ell \mathbf{y}_\ell + \mathbf{u}_\ell \right] \bmod \Lambda_F$ is independent of $\mathbf{y}_\ell$ and uniformly distributed over $\mathcal{V}(\Lambda_F)$ by the Crypto Lemma, $(a)$ holds from the distributive law and $(b)$ holds with high probability if $\beta_\ell^2 \left( P \|\mathbf{H}_{\ell,\mathcal{K}}\|^2+1 \right) + d_t < \theta_C$ which holds by choosing $\beta_\ell$ as in \eqref{SUC_op_2}.

Defining $$ \mathbf{t}_\ell \triangleq \widetilde{\mathbf{y}}_\ell/\beta_\ell = \left\{
	\begin{array}{ll}
		\left[  \mathbf{y}_\ell +\mathbf{q}_\ell \right] \bmod \Lambda_C  & \mbox{if } d_{\text{SU},\ell} > d_t \\
		  \mathbf{y}_\ell +\mathbf{q}_\ell  & \mbox{if } d_{\text{SU},\ell} < d_t
	\end{array}
\right.  $$ where $\mathbf{q}_\ell \triangleq \widetilde{\mathbf{q}}_\ell/\beta_\ell$ and $\frac{1}{T} \mathbb{E}\|\mathbf{q}_\ell\|^2=d_t/\beta_\ell^2$, the CP then forms linear combinations 
\begin{align}\label{s_estimate}
\widehat{\mathbf{v}}_{m} &= \left[ \sum\limits_{\ell=1}^{L} {a}_{m,\ell} \mathbf{t}_\ell \right] \bmod{\Lambda_C}\\ & \stackrel{(a)}{=} \left[\sum\limits_{\ell=1}^{L} {a}_{m,\ell} \left({\mathbf{y}}_\ell+\mathbf{q}_\ell \right) \right] \bmod{\Lambda_C} \\ &\stackrel{\text{(b)}}{=}  \sum\limits_{\ell=1}^{L} {a}_{s,m,\ell} \left({\mathbf{y}}_\ell +\mathbf{q}_\ell \right)
\end{align} where $(a)$ holds from the distributive law and $(b)$ holds w.h.p. if $$ \frac{1}{T} \mathbb{E} \|{\mathbf{v}}_{m}\|^2 = \mathbf{a}_{m}^\T \left( \mathbf{K}_{YY}+\mathbf{D}\right) \mathbf{a}_{m} < \theta_C,$$ where $\mathbf{D}=\text{diag}(d_1,\ldots,d_L)$ is the effective covariance matrix of $\mathbf{Q}$ and ${d}_\ell=d_t/\beta_\ell^2$ for $\ell \in \mathcal{L}$.

To guarantee correct recovery with probability at least $1-\rho_s$, $d_t$ should be chosen such that
\begin{align}
\mathbb{P}\Big(\max_m \mathbf{a}_{m}^\T \left( \mathbf{K}_{YY}+\mathbf{D} \right) \mathbf{a}_{m} \geq \theta_C \Big) = \rho_s \ .
\end{align} Finally, assuming correct recovery, the CP applies $\mathbf{A}^{-1}$ to obtain $ \widehat{\mathbf{Y}}=\mathbf{A}^{-1} \mathbf{V} = \mathbf{Y}+\mathbf{Q}$. 

\begin{lemma}
The symmetric compression rate for opportunistic IFSC is
\begin{align}\label{IFSC_rate_op}
 &R^s_{\text{IFSC,op}} (\mathbf{H},d_t) = \\ &\min\limits_{\substack{\mathbf{A} \in \mathbb{Z}^{L \times L} \\ \text{Rank}(\mathbf{A}) = L}} \max_\ell   \frac{1}{2} \log^+ \left( \frac{\mathbf{a}_{\ell}^\T \left( \mathbf{K}_{YY}+\mathbf{D}\right) \mathbf{a}_{\ell} }{d_t}  \right)\nonumber
\end{align} 
where 
\begin{align}
&\mathbf{D}=\text{diag}(d_t/\beta_1^2,\ldots,d_t/\beta_L^2) \label{D_for_opportunistic_IFSC} \\ 
 &\beta_\ell= \begin{dcases} 
		1 & \mbox{if } d_t \leq \frac{P\|\mathbf{H}_{\ell,\mathcal{K}}\|^2+1}{2^{2\C}-1} \\
		\sqrt{\frac{2^{2\C}-1}{P\|\mathbf{H}_{\ell,\mathcal{K}}\|^2+1} d_t}  & \mbox{if } d_t >\frac{P\|\mathbf{H}_{\ell,\mathcal{K}}\|^2+1}{2^{2\C}-1}
	\end{dcases}\label{beta_for_opportunistic_IFSC}
\end{align}
 and $d_t$ is chosen such that $\mathbb{P}(R^s_{\text{IFSC,op}} (\mathbf{H},d_t) > \C) = \rho_s$.
\end{lemma}

\section{Central Processor Channel Decoding}\label{sec: channel coding}
Once it has recovered the quantized observations $ \widehat{\mathbf{y}}_1 ,\ldots, \widehat{\mathbf{y}}_L$, the CP can act as the receiver in a virtual MAC to decode the transmitted codewords $\mathbf{x}_1 ,\ldots, \mathbf{x}_K$. It will be useful to write the recovered observations at the CP as
  \begin{align}
  \widehat{\mathbf{Y}} &=\mathbf{Y}+\mathbf{Q} =\mathbf{H}\mathbf{X}+\mathbf{Z}+\mathbf{Q}
 \end{align} where $ \mathbf{X} \triangleq [\mathbf{x}_1 ~\cdots~ \mathbf{x}_K]^\T  $, $\mathbf{Y} \triangleq [\mathbf{y}_1 ~\cdots~ \mathbf{y}_L]^\T$, $\widehat{\mathbf{Y}} \triangleq [\widehat{\mathbf{y}}_1 ~\cdots~ \widehat{\mathbf{y}}_L]^\T$,  $\mathbf{Z} \triangleq [\mathbf{z}(1) ~\cdots~ \mathbf{z}(T)]$ and $\mathbf{Q} \triangleq [\mathbf{q}_1 ~\cdots~ \mathbf{q}_L]^\T$ has a diagonal effective covariance matrix $ \mathbf{D} \triangleq \frac{1}{T} \mathbb{E}\left(\mathbf{Q}\mathbf{Q}^\T\right)$. Below, we present several decoding strategies and their achievable symmetric rates for a given channel realization $\mathbf{H}$ and distortion matrix $\mathbf{D}$. This then yields the symmetric rate as a function of $\mathbf{H}$, which can be plugged into Definition 1  to determine the outage rate. For the uplink C-RAN, the distortion levels in $\mathbf{D}$ are determined by the compression schemes chosen amongst those in Section \ref{sec: conventional compression}.

	\subsection{Conventional Decoders}

\subsubsection{Joint ML Decoding}
The best performance is attained by simultaneously decoding all codewords $\mathbf{x}_1,\ldots,\mathbf{x}_K$ via a joint maximum likelihood (ML) decoder. Although its implementation complexity scales exponentially with number of users $K$, we include it as a benchmark.

\lemma\label{ML} 
For a given channel matrix $\mathbf{H}$ and distortion matrix $\mathbf{D}$, the achievable symmetric rate using joint ML decoding is
 \begin{align}\label{ML_rate}
  R_{\text{ML}} (\mathbf{H},\mathbf{D}) = \min\limits_{\mathcal{S}\subseteq \mathcal{K}} \frac{1}{2 |\mathcal{S}|} \log \left( \frac{\left| P \mathbf{H}_{\mathcal{L},\mathcal{S}} \mathbf{H}_{\mathcal{L},\mathcal{S}}^\T + \mathbf{I}+ \mathbf{D}  \right|}{|\mathbf{I+D}|} \right).
 \end{align}
 Lemma \ref{ML} follows from using joint typicality analysis and can be considered a special case of \cite[Proposition 1]{zy14}.

{\remark[Local CSIR]
Under local CSIR, the channel outage probability constraint is reduced to half its value under global CSIR (i.e.,  $\frac{\rho}{2}$ instead of $\rho$), since the other half is reserved for the decompression outage event.}
 
\subsubsection{Single-User Decoding}
Since the complexity of joint ML decoding scales exponentially with the number of users, it is often of interest to find suboptimal decoding algorithms of lower complexity. For instance, the CP can apply a linear equalizer $\mathbf{B}$ to its reconstructed observations to get $\widetilde{\mathbf{Y}} = \mathbf{B} \mathbf{\widehat{Y}}$ and then apply a single-user decoder to each row of $\widetilde{\mathbf{Y}}$ to recover the individual codewords. Thus, each row of $\mathbf{B}$ should be selected to maximize the SINR for the desired codeword, which corresponds to the MMSE equalization vector. 

\begin{lemma}\label{MMSE}
For a given channel matrix $\mathbf{H}$ and distortion matrix $\mathbf{D}$, the achievable symmetric rate using an MMSE linear receiver is 
\begin{align}\label{MMSE_rate}
&R_\text{MMSE} (\mathbf{H},\mathbf{D}) =\\ & \frac{1}{2} \min\limits_{k \in \mathcal{K}} \log \left(1+ \frac{P (\mathbf{b}_{k}^\T \mathbf{H}_{\mathcal{L},k})^2}{ \mathbf{b}_{k}^\T (\mathbf{I+D})\mathbf{b}_k + P \sum\limits_{i \neq k} (\mathbf{b}_{k}^\T \mathbf{H}_{\mathcal{L},i})^2} \right) \nonumber
 \end{align}
where $\mathbf{b}_{k}^\T= P \mathbf{H}_{\mathcal{L},k}^\T \left(P \sum\limits_{j=1}^K \mathbf{H}_{\mathcal{L},j} \mathbf{H}_{\mathcal{L},j}^\T +\mathbf{I+D} \right)^{-1} $ is the $\lth$ row of the MMSE equalization matrix $\mathbf{B}$ and $\mathbf{H}_{\mathcal{L},k}$ is the $\kth$ column of the channel matrix $\mathbf{H}$. 
\end{lemma} 
See \cite[Section 8.3.3]{tse_book} for more details on MMSE decoders.
\subsubsection{Successive Interference Cancellation}
As in Wyner-Ziv compression, we can use recovered codewords as side information. In order to improve the performance, consider a decoding order defined by the permutation $\pi_c : \mathcal{K} \rightarrow \mathcal{K}$. The MMSE decoder with successive interference cancellation (MMSE-SIC) cancels out the effect of previously decoded codewords $\mathbf{x}_{\pi_c(1)}, \ldots, \mathbf{x}_{\pi_c(k-1)}$ (assuming successful decoding) before decoding the current codeword $\mathbf{x}_{\pi_c(k)}$, and then equalizes the result to get
\begin{align}
\widetilde{\mathbf{y}}_k^\T = \mathbf{b}_k^\T \left( \widehat{\mathbf{Y}} - \sum\limits_{i=1}^{k-1} \mathbf{H}_{\mathcal{L},\pi_c(i)} \mathbf{x}_{\pi_c(i)}^\T \right),
\end{align} which is subsequently fed to a single-user decoder to recover $\mathbf{x}_{\pi_c(k)}$.

\begin{lemma}\label{MMSE_SIC}
For a given channel matrix $\mathbf{H}$ and distortion matrix $\mathbf{D}$, the achievable symmetric rate using an MMSE-SIC decoder is 
 \begin{align}\label{MMSE_SIC_rate}
  &R_\text{MMSE-SIC} (\mathbf{H},\mathbf{D}) =\\ \nonumber &  \frac{1}{2} \max_{\pi_c} \min\limits_{k\in \mathcal{K}} \log \left( 1+\frac{P (\mathbf{b}_{k}^\T \mathbf{H}_{\mathcal{L},\pi_c(k)})^2}{ \mathbf{b}_{k}^\T (\mathbf{I+D})\mathbf{b}_k + P \sum\limits_{i > k} (\mathbf{b}_{k}^\T \mathbf{H}_{\mathcal{L},\pi_c(i)})^2} \right)
 \end{align}
where $\mathbf{b}_k^\T=P \mathbf{H}_{\mathcal{L},\pi_c(k)}^\T (P \sum\limits_{j \geq k} \mathbf{H}_{\mathcal{L},\pi_c(j)} \mathbf{H}_{\mathcal{L},\pi_c(j)}^\T+\mathbf{I}+ \mathbf{D} )^{-1}$ is the MMSE-SIC equalization vector. See \cite[Section 8.3.3]{tse_book} for more details on MMSE-SIC decoders.
\end{lemma}

\subsection{Integer-Forcing Decoding}\label{IFCC}
The idea behind an integer-forcing receiver is to switch the usual order of eliminating interference and denoising. It first, decodes integer-linear combinations of the transmitted codewords, and then solves for the desired codewords. Specifically, in order to decode the combinations $$ \mathbf{v}_{c,m}^\T \triangleq \mathbf{a}_{c,m}^\T \mathbf{X}, \ \ \forall m \in \mathcal{K}$$ where $\mathbf{a}_{c,m} \in \mathbb{Z}^K$, the CP first applies linear equalizers $\mathbf{b}_{c,m}^\T$ to get effective channels
\begin{align}
\widetilde{\mathbf{y}}_m^\T &= \mathbf{b}_{c,m}^\T \widehat{\mathbf{Y}} \nonumber \\ &= \underbrace{\mathbf{a}_{c,m}^\T \ \mathbf{X}}_\text{lattice codeword}+ \underbrace{(\mathbf{b}_{c,m}^\T \mathbf{H} -\mathbf{a}_{c,m}^\T) \mathbf{X} + \mathbf{b}_{c,m}^\T (\mathbf{Z}+\mathbf{Q})}_\text{effective noise} \\ &= \mathbf{v}_{c,m}^\T + \mathbf{z}_{\text{eff},m}^\T, \ \ \forall m \in \mathcal{K}
\end{align}
where $\mathbf{z}_{\text{eff},m}^\T= (\mathbf{b}_{c,m}^\T \mathbf{H} -\mathbf{a}_{c,m}^\T )\mathbf{X} +\mathbf{b}_{c,m}^\T (\mathbf{Z}+\mathbf{Q}) $ is the effective noise  due to the scaled AWGN $\mathbf{b}_{c,m}^\T \mathbf{Z}$, the scaled quantization noise $\mathbf{b}_{c,m}^\T \mathbf{Q}$ and the mismatch between the equalized channel $\mathbf{b}_{c,m}^\T\mathbf{H}$ and the integer vector $\mathbf{a}^\T_{c,m}$.
The CP then employs single-user decoders to decode $\mathbf{v}_{c,1},\ldots, \mathbf{v}_{c,K}$, and finally solves for $\mathbf{x}_1 ,\ldots, \mathbf{x}_K$.

The effective variance of $\mathbf{z}_{\text{eff},m}$ is
\begin{align}\label{eff_noise_var}
\sigma_{\text{eff},m}^2 & \triangleq \frac{1}{T} \mathbb{E}\|\mathbf{z}_{\text{eff},m}\|^2 \\ &=\|\mathbf{b}_{c,m}^\T \mathbf{H} - \mathbf{a}_{c,m}^\T\|^2 P + \mathbf{b}_{c,m}^\T \left( \mathbf{I}+\mathbf{D} \right) \mathbf{b}_{c,m} 
\end{align} 
where $\mathbf{D} \triangleq \text{diag} (d_1,\ldots,d_{L})$ is the covariance matrix of the quantization noise $\mathbf{Q}$.

Using the MMSE equalizer that minimizes the noise variance in \eqref{eff_noise_var}
$$ \mathbf{b}_{c,m}^\T = P \mathbf{a}_{c,m}^\T \mathbf{H}^\T  \left( P \mathbf{H} \mathbf{H}^{\T} + \mathbf{I} + \mathbf{D} \right)^{-1},$$
and applying Woodbury's matrix identity, we can write \eqref{eff_noise_var} as
\begin{align}
\sigma_{\text{eff},m}^2 &= \mathbf{a}_{c,m}^\T \left( P^{-1} \mathbf{I}+ \mathbf{H}^{\T}  \left( \mathbf{I +D} \right)^{-1}  \mathbf{H} \right)^{-1} \mathbf{a}_{c,m}\\  &= \| \mathbf{F}_c\,  \mathbf{a}_{c,m}\|^2
\end{align}
where $\mathbf{F}_c$ is any matrix satisfies the decomposition  $\mathbf{F}_c^{\T} \mathbf{F}_c =\left( P^{-1} \mathbf{I}+ \mathbf{H}^{\T} \left( \mathbf{I +D} \right)^{-1}  \mathbf{H} \right)^{-1} $.

\begin{lemma}
For a given channel matrix $\mathbf{H}$ and distortion matrix $\mathbf{D}$, the achievable symmetric rate for the integer-forcing strategy with parallel channel decoding is
\begin{align}\label{IFCC_rate}
R_\text{IFCC}(\mathbf{H},\mathbf{D}) = &\max\limits_{\substack{\mathbf{A}_c \in \mathbb{Z}^{K \times K} \\ \mathrm{rank}(\mathbf{A}_c) = K}} \min\limits_{m \in \mathcal{K}}  \frac{1}{2} \log^+ \left( \frac{P}{\|\mathbf{F}_c\, \mathbf{a}_{c,m} \|^2}\right).
\end{align}
\end{lemma}

{\remark Similar to MMSE-SIC, successive decoding for the combinations $\mathbf{v}_{c,1},\ldots,\mathbf{v}_{c,K}$ is possible and improves the achievable symmetric rate for IF receivers on average. See \cite{oen13} for more details.
}
\section{Integer-Forcing C-RAN Architecture}\label{sec: IF CRAN architecture}

The end-to-end integer-forcing architecture for C-RAN is illustrated in Figure~\ref{IFCRAN_Architecture}. It employs one of the integer-forcing source coding schemes in Section \ref{sec: lattice compression} to convey the channel observations to the CP, which then recovers the transmitted messages via integer-forcing channel decoding discussed in Section \ref{IFCC}.

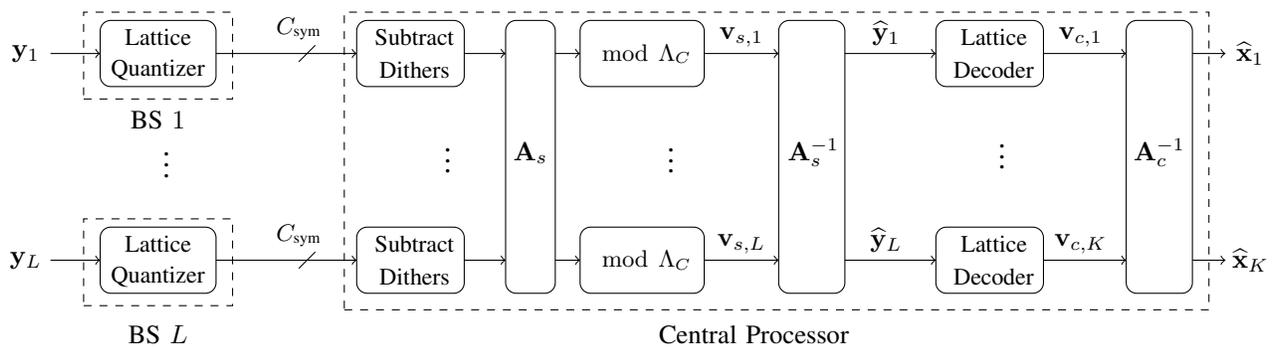
\begin{figure*}[h]
\centering
\captionsetup{justification=centering,margin=1cm}
\centering
{\begin{tikzpicture}[scale=1.1]

\node at (-0.3,1) { ${\mathbf{y}}_1$};
\node at (-0.3,-1.5) { ${\mathbf{y}}_L $};

\draw [->] (0,1) -- (.6,1);
\draw [->] (0,-1.5) -- (.6,-1.5);

\node at (1.4,-0.25) {\large\vdots};

\draw [black,rounded corners] (0.6,0.6) rectangle (2,1.4);
\draw [black,rounded corners] (0.6,-1.9) rectangle (2,-1.1);

\node at (1.3,1.2) {\small Lattice};
\node at (1.3,0.8) {\small Quantizer};

\node at (1.3,-1.3) {\small Lattice};
\node at (1.3,-1.7) {\small Quantizer};

\draw [->] (2,1) -- (3.7,1);
\draw [->] (2,-1.5) -- (3.7,-1.5);

\node at (4.8,-0.2) {\large$\vdots$};

\draw [-] (3,0.9)--(3.2,1.1);
\draw [-] (3,-1.6)--(3.2,-1.4);

\node at (3,1.3) {\small{$C_\text{sym}$}};
\node at (3,-1.2) {\small{$C_\text{sym}$}};

\draw [black,rounded corners] (3.7,0.6) rectangle (5,1.4);
\draw [black,rounded corners] (3.7,-1.9) rectangle (5,-1.1);

\draw [black,dashed] (.4,0.45) rectangle (2.2,1.5);
\draw [black,dashed] (0.4,-2.05) rectangle (2.2,-1);

\node at (1.3,0.2) {BS $1$};
\node at (1.3,-2.4) {BS $L$};

\node at (4.4,1.2) {\small Subtract};
\node at (4.4,0.8) {\small Dithers};

\node at (4.4,-1.3) {\small Subtract};
\node at (4.4,-1.7) {\small Dithers};

\draw [->] (5,1) -- (5.5,1);
\draw [->] (5,-1.5) -- (5.5,-1.5);

\draw [black,rounded corners] (5.5,-1.9) rectangle (6.1,1.4);
\node at (5.8,-0.2) {{$\mathbf{A}_s$}};

\draw [->] (6.1,1) -- (6.4,1);
\draw [->] (6.1,-1.5) -- (6.4,-1.5);

\draw [black,rounded corners] (6.4,0.6) rectangle (7.9,1.4);
\draw [black,rounded corners] (6.4,-1.9) rectangle (7.9,-1.1);
\node at (7.5,-0.2) {\large\vdots};

\node at (7.2,1) {\small $\bmod ~\Lambda_C$};
\node at (7.2,-1.5) {\small $\bmod ~\Lambda_C$};

\node at (8.35,1.2) { $\mathbf{v}_{s,1}$};
\node at (8.35,-1.3) { $\mathbf{v}_{s,L} $};

\draw [->] (7.9,1) -- (8.8,1);
\draw [->] (7.9,-1.5) -- (8.8,-1.5);

\draw [black,rounded corners] (8.8,-1.9) rectangle (9.6,1.4);
\node at (9.22,-0.2) {{$\mathbf{A}_s^{-1}$}};

\node at (10.1,1.25) { $\widehat{\mathbf{y}}_1$};
\node at (10.1,-1.25) { $\widehat{\mathbf{y}}_L $};

\draw [->] (9.6,1) -- (10.7,1);
\draw [->] (9.6,-1.5) -- (10.7,-1.5);

\draw [black,rounded corners] (10.7,0.6) rectangle (12,1.4);
\draw [black,rounded corners] (10.7,-1.9) rectangle (12,-1.1);
\node at (11.5,-0.15) {\large\vdots};

\node at (11.4,1.2) {\small Lattice};
\node at (11.4,0.8) {\small Decoder};

\node at (11.4,-1.3) {\small Lattice};
\node at (11.4,-1.7) {\small Decoder};

\draw [->] (12,1) -- (13,1);
\draw [->] (12,-1.5) -- (13,-1.5);

\node at (12.45,1.2) { $\mathbf{v}_{c,1}$};
\node at (12.45,-1.3) { $\mathbf{v}_{c,K} $};

\draw [black,rounded corners] (13,-1.9) rectangle (13.8,1.4);
\node at (13.4,-0.2) {{$\mathbf{A}_c^{-1}$}};

\draw [->] (13.8,1) -- (14.2,1);
\draw [->] (13.8,-1.5) -- (14.2,-1.5);

\node at (14.5,1) { $\widehat{\mathbf{x}}_1$};
\node at (14.5,-1.5) { $\widehat{\mathbf{x}}_K $};

\draw [black,dashed] (3.55,-2.1) rectangle (14,1.5);
\node at (8.5,-2.4) {{Central Processor}};
\end{tikzpicture}}
\caption{Integer-forcing architecture for C-RAN with symmetric distortion.}
\vspace{-0.15in}
\label{IFCRAN_Architecture}
\end{figure*}

\theorem\label{IF_CRAN_th_sym}
The achievable symmetric rate for the IF C-RAN strategy with global CSIR, parallel decompression, and parallel channel decoding is
\begin{equation}\begin{aligned}\label{IF_CRAN_rate_sym}
R_\text{IF-CRAN}(\mathbf{H}) = &\max_{d,\mathbf{A}_c \in \mathbb{Z}^{K \times K}}  \min\limits_{m \in \mathcal{K}}  \frac{1}{2} \log^+ \left( \frac{P}{\|\mathbf{F}_c \,  \mathbf{a}_{c,m} \|^2}\right)\\
&\text{~~~~subject to~~}   \text{Rank}(\mathbf{A}_c)=K\\ &~~~~~~~~~~~\text{and}~~ R^s_\text{IFSC} (\mathbf{H},d) \leq C_\text{sym}
 \end{aligned}\end{equation} where $R^s_\text{IFSC} (\mathbf{H},d)$ is from \eqref{IFSC_rate1} and $\mathbf{F}_c$ is any matrix satisfying $\mathbf{F}_c^{\T} \, \mathbf{F}_c =\left( P^{-1} \mathbf{I}+ \frac{1}{d+1} \mathbf{H}^{\T} \mathbf{H} \right)^{-1} $.  
Furthermore, the end-to-end performance can be enhanced by using asymmetric distortions for IFSC through algebraic successive decompression.

\theorem\label{IF_CRAN_th_asym}
The achievable symmetric rate for the IF C-RAN strategy with global CSIR, algebraic SIC decompression, and parallel channel decoding is
\begin{align}\label{IF_CRAN_rate_asym}
R_\text{IF-CRAN}(\mathbf{H}) = &\max\limits_{\mathbf{D},\mathbf{A}_c \in \mathbb{Z}^{K \times K}}  \min\limits_{m \in \mathcal{K}}  \frac{1}{2} \log^+ \left( \frac{P}{\|\mathbf{F}_c \, \mathbf{a}_{c,m} \|^2}\right)  \\
&\text{~~~~subject to~~}    \text{Rank}(\mathbf{A}_c)=K \\ & \text{~~~~~~~~~~~and~~}  R^s_{\text{IFSC},\ell} (\mathbf{H},\mathbf{D}) \leq C_\text{sym},~~  \forall \ell \in \mathcal{L} 
\end{align} where $ R^s_{\text{IFSC},\ell} (\mathbf{H},\mathbf{D})$ is from \eqref{IFSC_rate_SIC} and $\mathbf{F}_c$ satisfies  $\mathbf{F}_c^{\T} \mathbf{F}_c =\left( P^{-1} \mathbf{I}+ \mathbf{H}^{\T}  \left( \mathbf{I +D} \right)^{-1} \mathbf{H} \right)^{-1} $.

\theorem\label{Op_IF_CRAN}
The achievable symmetric rate for the IF C-RAN strategy with local CSIR, opportunistic IFSC, and parallel channel decoding is
\begin{equation}\label{IF_CRAN_rate_op}
\begin{aligned}
R_\text{IF-CRAN}(\mathbf{H}) = \max\limits_{d_t,\mathbf{A}_c \in \mathbb{Z}^{K \times K}} &  \min\limits_{m \in \mathcal{K}}  \frac{1}{2} \log^+ \left( \frac{P}{\|\mathbf{F}_c \, \mathbf{a}_{c,m} \|^2}\right)  \\
\text{subject to~~} & \text{Rank}(\mathbf{A}_c)=K  
\end{aligned}
\end{equation}

where $\mathbf{F}_c$ is any matrix that satisfies $\mathbf{F}_c^{\T} \mathbf{F}_c =\left( P^{-1} \mathbf{I}+ \mathbf{H}^{\T}  \left( \mathbf{I +D} \right)^{-1} \mathbf{H} \right)^{-1} $, $\mathbf{D}=\text{diag}(d_t/\beta_1^2,\ldots,d_t/\beta_L^2)$, $\beta_\ell$ is given by \eqref{beta_for_opportunistic_IFSC} and $d_t$ is chosen such that $\mathbb{P}\left( R^s_\text{IFSC,op}(\mathbf{H},d_t) > \C  \right) = \rho_s$ where $ R^s_\text{IFSC,op}(\mathbf{H},d_t)$ is given by \eqref{IFSC_rate_op} and $\rho_s$ is the compression outage probability.

The optimization problems in Theorems~\ref{IF_CRAN_th_sym}, \ref{IF_CRAN_th_asym}, and \ref{Op_IF_CRAN} are quite challenging due to the maximization over full-rank integer matrices as well as the non-convex objective of selecting the distortion levels. We propose (sub-optimal) algorithms for choosing the integer matrices as well the distortion levels in Section~\ref{sec: IF CRAN algorithms}.

\section{IF Outage Upper Bound}\label{sec: IF outage bound}
As noted in \cite{zneg14}, for some channel realizations $\mathbf{H}$, the achievable rate of IF channel coding can be far from the MIMO capacity. However,~\cite{de18} quantifies the measure of such channels $\mathbf{H}$ for the important special case of Gaussian fading (i.e., $\mathbf{H} \sim \mathcal{N}(\mathbf{0},\mathbf{I})$). A similar story holds for IF source coding as shown in~\cite{de17}: for certain covariance matrices of the form $P \mathbf{H}\mathbf{H}^{\T} + \mathbf{I}$, the performance falls short of BT compression, but, for i.i.d.~Gaussian~$\mathbf{H}$, the measure of such ``difficult'' channels can be bounded. Here, we combine ideas from the proofs in~\cite{de18,de17} to bound the measure of channels for which our IF-CRAN scheme falls significantly below the uplink C-RAN capacity.

To this end, we first express the IF rate in \eqref{IF_CRAN_rate_sym} in terms of the $K^{\text{th}}$ successive minima of the lattice $\mathbf{S}_2^{-\frac{1}{2}} \mathbf{U}^\T \mathbb{Z}^{K}$ where the diagonal matrix $\mathbf{S}_2$ and the orthogonal matrix $\mathbf{U}$ stem from the eigenvalue decomposition $\mathbf{U}\mathbf{S}_2\mathbf{U}^\T =\frac{P}{d+1} \mathbf{H}^\T \mathbf{H}+\mathbf{I}$. We then lower bound this rate expression using Lemma~\ref{thm:transference}:
\begin{align}
R_\text{IF-CRAN}(\mathbf{H}) &=  \frac{1}{2} \log \Bigg(\frac{1}{ {\lambda}_K^2 \big( \mathbf{S}_2^{-\frac{1}{2}}\mathbf{U}^\T  \big)} \Bigg) \\ &\geq  \frac{1}{2} \log \Bigg( \frac{\lambda_1^2 \big( \mathbf{S}_2^{\frac{1}{2}}\mathbf{U}^\T  \big)}{K^2} \Bigg) \\&=\frac{1}{2} \log \Bigg( \min_{\substack{\mathbf{a} \in \mathbb{Z}^{K} \\ \mathbf{a} \neq \mathbf{0}}} \frac{\|\mathbf{S}_2^{\frac{1}{2}}\mathbf{U}^\T \mathbf{a}\|^2 }{K^2} \Bigg). \label{IF_rate_in_terms_of_successive_minima}
\end{align} 
 
We now recall a result from~\cite{de18} that provides a bound on the outage probability for integer-forcing over i.i.d.~Gaussian fading. We make a slight modification to the original proof by using the Banaszczyk transference theorem from Lemma~\ref{thm:transference} to exchange $\alpha(K)$ in~\cite[Equation 36]{de18} with $K^2$, which yields the following lemma, whose form is more convenient for our analysis.
 
\lemma[{{\cite[Theorem 1]{de18}}}] \label{successive minima bound} For the Gaussian MAC (i.e., $C_\text{sym}=\infty$ and $d=0$) and any constant $\Delta C_\text{MAC}>0$, we have 
\begin{align}
\mathbb{P}\left( \min_{\substack{\mathbf{a}\in \mathbb{Z}^K \\ \mathbf{a} \neq  \mathbf{0}}} \|\mathbf{S}_1^{\frac{1}{2}} \mathbf{U}^\T \mathbf{a}\|^2 < 2^{\frac{2(C- \Delta C_\text{MAC})}{K}} K^2 \right) \leq \gamma(K) 2^{-\Delta C_\text{MAC}}
\end{align} where the orthogonal matrix $\mathbf{U}$ and the diagonal matrix $\mathbf{S}_1$ comes from the eigenvalue decomposition $\mathbf{U}\mathbf{S}_1\mathbf{U}^\T =P \mathbf{H}^\T \mathbf{H}+\mathbf{I}$, $C=\frac{1}{2} \log |\mathbf{S}_1|$ is the MAC capacity and $\gamma(K)$ is defined in \cite[Equation (59)]{de18} as $c(K)$ with replacing $\alpha(K)$ by $K^2$.

Let us define the probability that the difference between the IF C-RAN achievable rate and a cut-set bound on the sum capacity is larger than some positive constant $\dc$ as $$ 
P_\text{diff}(\Delta C) \triangleq  \mathbb{P} \left( K R_\text{IF-CRAN} (\mathbf{H}) < C_\text{upper}(\mathbf{H}) - \dc \right) $$ where $\Delta C>0$ is some constant and $C_\text{upper}(\mathbf{H}) \triangleq \min \left\{ L \C, \frac{1}{2} \log \left( \big| P\mathbf{H}^\T \mathbf{H} + \mathbf{I} \big| \right) \right\}$ is a cut-set bound for the sum capacity and the probability is taken with respect to $\mathbf{H}\sim \mathcal{N}(\mathbf{0},\mathbf{I})$.

\lemma \label{thm:diffprob}
For the uplink C-RAN channel with i.i.d.~Gaussian fading, $\mathbf{H}\sim \mathcal{N}(\mathbf{0},\mathbf{I})$, the probability that the rate of the integer-forcing strategy with global CSIR, parallel decompression, and parallel channel decoding is not within $\Delta C$ of the sum-capacity is upper bounded by
\begin{align}
P_\text{diff}(\Delta C) \leq \gamma(\max\{K,L\}) \ 2^{-\dc/3} 
\end{align} where $\gamma(\max\{K,L\})$ is defined in \cite[Equation (45)]{de17} as $c(\max\{K,L\})$.

$\textit{Proof}$: The proof closely follows that of \cite[Theorem 1]{de18}. We start by bounding $P_\text{diff}(\Delta C)$ as
\begin{align}
&P_\text{diff}(\Delta C) 
\\ &\leq \underbrace{\mathbb{P} \left( K R_\text{IF-CRAN}(\mathbf{H}) < C_\text{upper}(\mathbf{H}) - \dc \big| \mathcal{A} \right)}_{{(i)}} +\underbrace{\mathbb{P} \left( \mathcal{A}^c \right)}_{(ii)}.
\end{align} where $\mathcal{A} \triangleq  \{ R_\text{IFSC}^s(\mathbf{H},d^*) <  R^s_\text{BT}(\mathbf{H},d^*)+ \dr  \}$ is the event that the IFSC rate is within a constant $\Delta R >0$ (to be chosen later) from the BT compression rate and $d^*>0$ is the distortion that saturates the fronthaul rate constraint $R^s_\text{IFSC}(\mathbf{H},d^*)=\C$. For the rest of the proof, we will omit $d^*$ from $R^s_\text{IFSC}(\mathbf{H},d^*)$ and $R^s_\text{BT}(\mathbf{H},d^*)$ for the sake of conciseness.

Using \cite[Theorem 1]{de17}, we immediately have the upper bound $(ii)\leq \gamma(L) 2^{-\dr}$ where $\gamma(L)$ is defined in \cite[Equation (45)]{de17}. Next, to bound $(i)$, we use \eqref{IF_rate_in_terms_of_successive_minima} to get
\begin{align}
(i) &\stackrel{}{\leq} \mathbb{P}_{\mathbf{U},\mathbf{S}_1} \left( \min_{\mathbf{a}} \|\mathbf{S}_2^{1/2}\mathbf{U}^\T \mathbf{a}\|^2 < K^2 2^{2(C_\text{upper}(\mathbf{H})-\dc)/K} \big| \mathcal{A} \right) \nonumber \\
&= \mathbb{P}_{\mathbf{S}_1} \bigg[ \mathbb{P}_{\mathbf{U}|\mathbf{S}_1} \bigg( \min_{\mathbf{a}} \|\mathbf{S}_2^{1/2}\mathbf{U}^\T \mathbf{a}\|^2 \\ &\qquad~~~~    < K^2 2^{-2\dc/K} \min \{|\mathbf{S}_1|^{1/K},2^{2L\C/K}\} \big| \mathcal{A} \bigg) \bigg] \\
&={\mathbb{P}_{\mathbf{S}_1} \bigg[ \mathbb{P}_{\mathbf{U}|\mathbf{S}_1} \bigg( \min_{\mathbf{a}} \|\widetilde{\mathbf{S}}_2^{1/2} \mathbf{U}^\T \mathbf{a}\|^2}  \label{innerterm}\\ & \quad  {<K^2 2^{-2\dc/K} \min \left\{ \left(\frac{|\mathbf{S}_1|}{|\mathbf{S}_2|} \right)^{1/K}, \frac{2^{2L\C/K}}{|\mathbf{S}_2|^{1/K}} \right\} \bigg| \mathcal{A}  \bigg) \bigg]}
\end{align} where the minimization is over all non-zero integer vectors $\mathbf{a} \in \mathbb{Z}^{K} \setminus \{ \mathbf{0} \}$, $\mathbf{U}$, $\mathbf{S}_1$ and $\mathbf{S}_2$ come from the eigenvalue decompositions $\mathbf{U}\mathbf{S}_1 \mathbf{U}^\T = P \mathbf{H}^\T\mathbf{H} + \mathbf{I}$ and $\mathbf{U}\mathbf{S}_2 \mathbf{U}^\T = \frac{P}{d^*+1} \mathbf{H}^\T\mathbf{H} + \mathbf{I}$, and $\widetilde{\mathbf{S}}_2 \triangleq \frac{\mathbf{S}_2}{|\mathbf{S}_2|^{1/K}}$.

We now proceed to bound the RHS of the inequality inside \eqref{innerterm} for any value of $\mathbf{S}_1$ assuming that event $\mathcal{A}$. 
\begin{align}\label{inner_prob}
&K^2 2^{-2\dc/K} \min \left\{ \left(\frac{|\mathbf{S}_1|}{|\mathbf{S}_2|} \right)^{1/K}, \frac{2^{2L\C/K}}{|\mathbf{S}_2|^{1/K}} \right\}  \nonumber \\
&\stackrel{(a)}{\leq} \scalemath{0.9}{K^2 2^{-2\dc/K} \min\left\{ (d^*+1)^{L/K} , \frac{ 2^{2L\C/K}}{2^{2R^s_\text{BT}(\mathbf{H})/K} } \left(\frac{d^*+1}{d^*}\right)^{L/K} \right\}}  \nonumber \\
&\stackrel{(b)}{\leq} \scalemath{0.9}{K^2 2^{-2\dc/K} \min\left\{ (d^*+1)^{L/K} , 2^{2\dr/K} \left(\frac{d^*+1}{d^*}\right)^{L/K} \right\} }
\end{align} where $(a)$  holds from $|\mathbf{S}_2| =|\frac{1}{d^*+1} (\mathbf{S}_1+d^* \mathbf{I})| >  |\frac{1}{d^*+1}\mathbf{S}_1| $ and $R^s_\text{BT}(\mathbf{H}) = \frac{1}{2} \log \left| \frac{1}{d^*} \mathbf{K}_{YY}+\mathbf{I} \right|= \frac{1}{2} \log \left| \frac{d^*+1}{d^*} \mathbf{S}_2 \right|$ and $(b)$ holds from $ R_\text{BT}^s(\mathbf{H}) \geq  R_\text{IFSC}^s(\mathbf{H})-\dr = L\C-\dr $ given $\mathcal{A}$.

Next, we partition the space of possible values of $\mathbf{S}_1$ into $\mathcal{B}$ and $\mathcal{B}^c$, where $\mathcal{B} \triangleq \left\{ \frac{1}{2} \log |\mathbf{S}_1| > L \C  -L/2 - \dr  \right\}$ and bound $d^*$ depending on the event $\mathcal{B}$ as in Lemma~\ref{d_bounds} in Appendix B. Using \eqref{inner_prob}, we can upper bound $(i)$ by 
\begin{align}\label{inner_prob2}
&\scalemath{0.75}{ \mathbb{P}_{\mathbf{S}_1} \left[ \mathbb{P}_{\mathbf{U}|\mathbf{S}_1} \left( \min_{\mathbf{a}} \|\widetilde{\mathbf{S}}_2^{1/2} \mathbf{U}^\T \mathbf{a}\|^2 < K^2 2^{-2(\dc-\dr)/K} \left(\frac{d^*+1}{d^*}\right)^{L/K} \bigg| \mathcal{A},\mathcal{B} \right) \mathds{1}_{ \mathcal{B}} \right]} \nonumber \\
&~~+ \scalemath{0.75}{\mathbb{P}_{\mathbf{S}_1} \left[ \mathbb{P}_{\mathbf{U}|\mathbf{S}_1} \left( \min_{\mathbf{a}} \|\widetilde{\mathbf{S}}_2^{1/2} \mathbf{U}^\T \mathbf{a}\|^2 < K^2 2^{-2\dc/K} (d^*+1)^{L/K} \bigg| \mathcal{A},\mathcal{B}^c \right)  \mathds{1}_{ \mathcal{B}^c} \right]} \nonumber \\
&\stackrel{(a)}{\leq} \scalemath{0.75}{ \mathbb{P}_{\mathbf{S}_1} \left[ \mathbb{P}_{\mathbf{U}|\mathbf{S}_1} \left( \min_{\mathbf{a}} \|\widetilde{\mathbf{S}}_2^{1/2} \mathbf{U}^\T \mathbf{a}\|^2 < K^2 2^{-2(\dc-\dr)/K} 2^{2(\dr+L)/K} \bigg| \mathcal{A},\mathcal{B}  \right) \mathds{1}_{\mathcal{B}} \right] } \nonumber \\
& ~~+ \scalemath{0.75}{ \mathbb{P}_{\mathbf{S}_1} \left[ \mathbb{P}_{\mathbf{U}|\mathbf{S}_1} \left( \min_{\mathbf{a}} \|\widetilde{\mathbf{S}}_2^{1/2} \mathbf{U}^\T\mathbf{a}\|^2 < K^2 2^{-2\dc/K} 2^{L/K} \bigg| \mathcal{A},\mathcal{B}^c  \right)  \mathds{1}_{\mathcal{B}^c} \right] }\nonumber \\
&\stackrel{(b)}{\leq} \gamma(K) 2^{-(\dc - 2\dr)} 2^{L} +\gamma(K) 2^{-(\dc)} 2^{L/2}  
\end{align}
where $\mathds{1}$ is the indicator function, $(a)$ holds from Lemma \ref{d_bounds} in Appendix B and $(b)$ holds from Lemma \ref{successive minima bound} by substituting $\Delta C_\text{MAC}= \Delta C - 2 \Delta R -L$ and $\Delta C_\text{MAC}= \Delta C - L/2$, respectively.

 The rest of the proof follows by combining $(i)$ and $(ii)$ and taking $\dr=\frac{\dc}{3}$ so that the exponential terms in $(ii)$ and \eqref{inner_prob2} are $\frac{\dc}{3}$.
 
For a fixed sum rate $R$, define the optimal outage probability as $p_\text{optimal}(R) \triangleq \mathbb{P} \left( C(\mathbf{H}) < R \right) $, where $C(\mathbf{H})$ is the sum capacity of the uplink C-RAN channel. The theorem below shows that IF is approximately optimal in the following sense: it can operate within a constant gap of the optimal tradeoff between outage rate and probability.   

\theorem For a positive constant $\Delta C$, the outage probability for the integer-forcing C-RAN strategy with global CSIR, parallel decompression, and parallel channel decoding is bounded by
\begin{align}
p_\text{IF-CRAN}(R-\Delta C) \leq p_\text{optimal}(R)+ \gamma(\max\{K,L\}) \ 2^{-\dc/3}.
\end{align} 

$\textit{Proof}$: Using the law of total probability, the IF-CRAN outage probability can be written as
\begin{align}
&p_\text{IF-CRAN}(R-\Delta C)\\ &= \mathbb{P}\left( \left\{ KR_\text{IF-CRAN} \leq R-\Delta C \right\} \cap \left\{ C(\mathbf{H}) \geq R \right\} \right) \nonumber \\ 
&~~+~\mathbb{P}\left( KR_\text{IF-CRAN} \leq R-\Delta C \big| C(\mathbf{H})<R \right) \mathbb{P} \left( C(\mathbf{H})<R \right) \nonumber \\
&\leq \mathbb{P}\left( KR_\text{IF-CRAN} \leq C(\mathbf{H})-\Delta C \right) + \mathbb{P} \left( C(\mathbf{H})<R \right) \nonumber \\
&\leq \mathbb{P}\left( KR_\text{IF-CRAN} \leq C_\text{upper}(\mathbf{H})-\Delta C \right) + \mathbb{P} \left( C(\mathbf{H})<R \right) \nonumber \\
&\leq  \gamma(\max\{K,L\}) \ 2^{-\dc/3}+p_\text{optimal}(R)
\end{align} where $C_\text{upper}(\mathbf{H}) \triangleq \min \left\{ L \C, \frac{1}{2} \log  \big| P\mathbf{H}^\T \mathbf{H} + \mathbf{I} \big|  \right\}$ is a cut-set bound on the sum capacity of the uplink C-RAN channel and we used Lemma~\ref{thm:diffprob} in the last step.

\remark Recall that, if the transmitters have enough CSIT to set their rates, then WZC with MMSE-SIC can operate within a constant gap of the sum capacity of any C-RAN channel~\cite{ZXYC16}. In contrast, we cannot give such a guarantee for IF, even with CSIT, since there are difficult channel realizations for which the gap can be arbitrarily large~\cite{zneg14}. Yet, the theorem shows that, these channels have small measure and that IF can operate within a constant gap of the optimal outage rate-probability curve (without any CSIT). 


\section{Optimization Algorithms} \label{sec: IF CRAN algorithms}
In this section, we propose algorithms that can be used to select the parameters of the IF-CRAN scheme proposed in Section \ref{sec: IF CRAN architecture}.

\subsection{IF-CRAN with Symmetric Distortion}
The optimization problems from Theorems \ref{IF_CRAN_th_sym}, \ref{IF_CRAN_th_asym}, and \ref{Op_IF_CRAN} are challenging due to the integer constraints on $\mathbf{A}_c$ and $\mathbf{A}_s$. Specifically, for a fixed distortion level $d$ the problems of finding the optimal integer matrix $\mathbf{A}_s$ to minimize the symmetric compression rate or finding the optimal integer matrix $\mathbf{A}_c$ to maximize the IF C-RAN symmetric transmission rate are linked to the hard combinatorial problem of finding the shortest set of linearly independent lattice vectors \cite{mg02book}.

For a fixed matrix $\mathbf{A}_c$, the overall rate in \eqref{IF_CRAN_rate_sym} is monotonically increasing in $d$. Using a bisection search, we can quickly converge to the smallest $d$ that meets the fronthaul constraint (i.e., $R_{\text{IFSC}}(\mathbf{H},d)=\C$). During each iteration in the search, $\mathbf{A}_s$ can be optimized using an LLL reduction \cite{lll82} on the induced lattice $\mathbf{F}_s$, which provides an approximate guarantees. A detailed algorithm is given in Algorithm \ref{alg_sym_IFSC}. See Figure~\ref{f:convergence} to see that $d$ converges within a few iterations. Finally, once we find a solution for $\mathbf{A}_s$ and $d$ that meets the fronthaul constraints using Algorithm \ref{alg_sym_IFSC}, an approximate solution for the integer matrix $\mathbf{A}_c$ can be obtained using an LLL reduction on the basis $\mathbf{F}_c$. 

\begin{algorithm}[htbp]
  \caption{Symmetric IFSC}\label{alg_sym_IFSC}
  \begin{algorithmic}[1]
    \Procedure{SIFSC}{P,$\mathbf{H},C_\text{sym}$,tol}
      \State Initialization: Set  $d_\text{min}=0$ and $d_\text{max}=d$ large enough such that $R_{\text{IFSC}} (\mathbf{H},d)<C_\text{sym}$.
      \While{$C_\text{sym}-R_{\text{IFSC}} (\mathbf{H},d)>\text{tol}$ \ \Or \ $R_{\text{IFSC}} (\mathbf{H},d)> C_\text{sym}$}\label{while_iter1}
    	 \If{$R_{\text{IFSC}} (\mathbf{H},d) < C_\text{sym}$}\label{for_iter}
	 	    \State $d_\text{max}=d/2.$
	     \Else
	 	    \State $d_\text{min}=d/2.$
	     \EndIf
   	     \State $d=(d_\text{min}+d_\text{max})/2,$
	     \State $\mathbf{F}_s= \text{chol}((1+\frac{1}{d})\mathbf{I}+ \frac{1}{d} P \mathbf{H}\mathbf{H}^\T)$
      \State $\mathbf{A}_{s}= \text{LLL-reduction} (\mathbf{F}_s)$, \State $R_{\text{IFSC}} (\mathbf{H},d) = \frac{1}{2} \log^+ (\|\mathbf{F}_s \mathbf{a}_{s,L}\|^2)$
      \EndWhile\label{while_iter2}
      \State \textbf{return} $d$.
    \EndProcedure
  \end{algorithmic}
\end{algorithm}

\subsection{IF-CRAN with Asymmetric Distortion}
In the symmetric case, we were able to decouple the problem of choosing the distortion level $d$ from the problem of choosing the integer matrix $\mathbf{A}_c$. However, in the case of asymmetric distortion levels in \eqref{IF_CRAN_rate_asym}, both problems are more tightly coupled. In order to tackle this problem, we initially set all distortion levels to the symmetric value $d$ such that $R_{\text{IFSC},\ell}(\mathbf{H},d\mathbf{I})=\C$, fix the integer matrix $\mathbf{A}_s$, then find the distortion levels $d_1,\ldots,d_L$ such that $R_{\text{IFSC},\ell}(\mathbf{H},\mathbf{D}) = \C, \forall \ell \in \mathcal{L}$. With $\mathbf{A}_s$ fixed, this corresponds to solving $L$ linear equations for $\mathbf{D}$. Finally, we need to permute the BSs before and after solving for $\mathbf{D}$ to obtain full-rank sub-matrices $\mathbf{A}_{s,[1:m]}$ for $m=1,\ldots,L$ so that we can use the rates in \eqref{IFSC_rate_SIC}. Details are given in Algorithm \ref{AIFSC}.

\begin{algorithm}
  \caption{Asymmetric IFSC}\label{AIFSC}
  \begin{algorithmic}[1]
    \Procedure{AIFSC}{$\mathbf{K}_{{Y}{Y}},C_\text{sym}$}
     
      \State Initialization: Fix $d_\ell=d, \forall \ell$ and solve $d=\text{SIFSC}(\mathbf{H},\C,\text{tol})$.

      \State Fix $\mathbf{A}_s$ and find permutation $\pi_\text{IF}$  s.t. rank($\mathbf{A}_{s,[1:m],\pi_\text{IF}([1:m])})=m, \  \forall m =1,\ldots,L$. 
      \State Find distortion levels $d_1,\ldots,d_L$ satisfying $\mathbf{C}  [d_1 \cdots d_L]^\T = \mathbf{e}$ where $\mathbf{C} \triangleq 2^{\C} \times \mathbf{I} -\mathbf{A}_{s,\mathcal{L},\pi_\text{IF}(\mathcal{L})} \odot \mathbf{A}_{s,\mathcal{L},\pi_\text{IF}(\mathcal{L})}$ and ${e}_\ell \triangleq \mathbf{a}_{s,\ell}^\T\mathbf{K}_{YY} \mathbf{a}_{s,\ell}, \forall \ell \in \mathcal{L}$.
     
      \State \textbf{return} $\mathbf{D} = \text{diag}(d_{\pi^{-1}_\text{IF}(1)},\ldots,d_{\pi^{-1}_\text{IF}(L)}) $.
    \EndProcedure
  \end{algorithmic}
\end{algorithm}

\remark
The asymmetric distortion levels obtained from Algorithm~\ref{AIFSC} are upper-bounded by the distortion level obtained from Algorithm~\ref{alg_sym_IFSC}. This is because the symmetric distortion $d$ that satisfies $ R_{\text{IFSC}}(\mathbf{H},d)=\C$ (i.e., Algorithm~\ref{AIFSC} result) also guarantees that $R_{\text{IFSC},\ell}(\mathbf{H},d\mathbf{I}) \leq \C, \forall \ell  \in \mathcal{L}$, since for both cases, the integer matrix $\mathbf{A}_s$ is the same and in IFSC with parallel decoding, all rates are constrained by the combination with the largest variance. Second, decreasing one distortion level only increases the compression rate of the corresponding BS and simultaneously decreases the rate of the other BSs.

\section{Simulations}\label{sec: simulation results}

We now turn to numerical evaluations of the IF and WZ rate expressions for both global and local CSIR in order to gain insights as to the performance differences between these two competing architectures. For each plot, we generated $1000$ independent realizations for the channel matrix $\mathbf{H}$, elementwise i.i.d.~$\mathcal{N}(0,1)$. Ideally, we would also plot the exact uplink C-RAN capacity expression as a benchmark. Since the exact capacity is an open problem, we instead use BT compression with symmetric distortion and joint ML decoding as a benchmark, which is known to attain the capacity within a constant gap \cite{gk17}.

    \begin{figure*}[!h]
\begin{subfigure}[t]{.49\linewidth}
    \centering\includegraphics[width=0.9\linewidth]{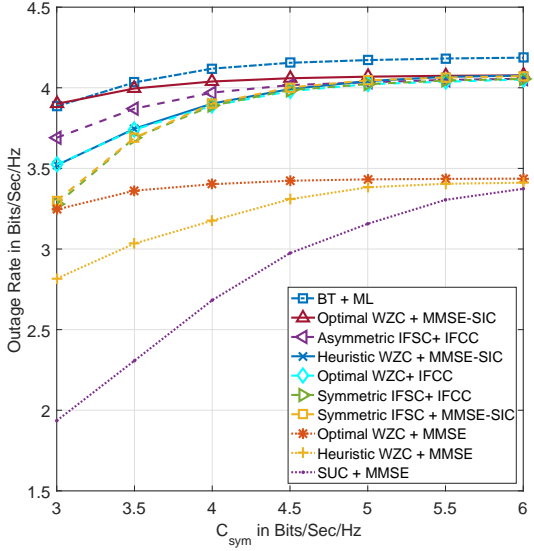}
    \caption{Outage rate per user.}\label{f:global_3_6}
  \end{subfigure}
  \begin{subfigure}[t]{.49\linewidth}
    \centering\includegraphics[width=0.9\linewidth]{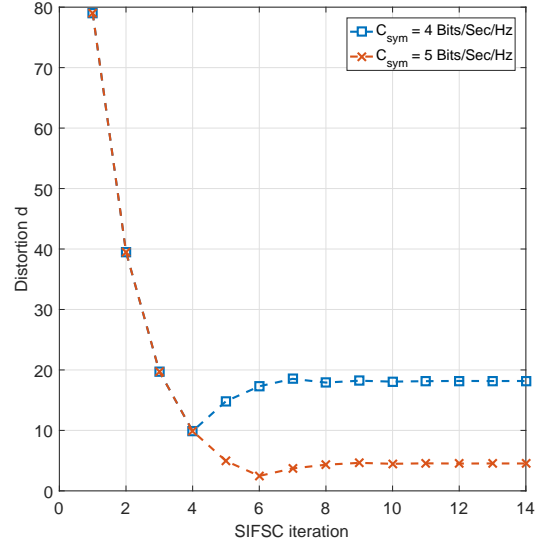}
        \caption{Convergence of SIFSC Algorithm.} \label{f:convergence}\quad\quad
  \end{subfigure}
  \vspace{-0.05in}
  \caption{Global CSIR with $K = 3$ users, $L = 6$ basestations, $\mathsf{SNR} = 25$dB, and $5\%$ outage. }\label{local_snr}\vspace{-0.2in}
\end{figure*}

\begin{figure*}[!h]
\begin{subfigure}[t]{.49\linewidth}
    \centering\includegraphics[width=0.9\linewidth]{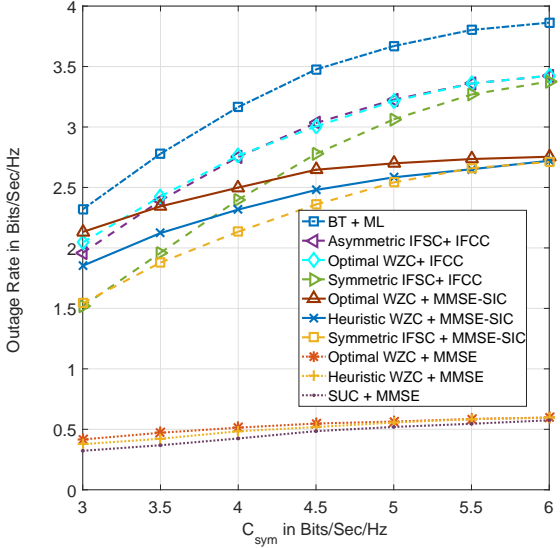}
        \caption{$L=6$ basestations.}\label{f:global_6_6}
  \end{subfigure}
  \begin{subfigure}[t]{.49\linewidth}
    \centering\includegraphics[width=0.9\linewidth]{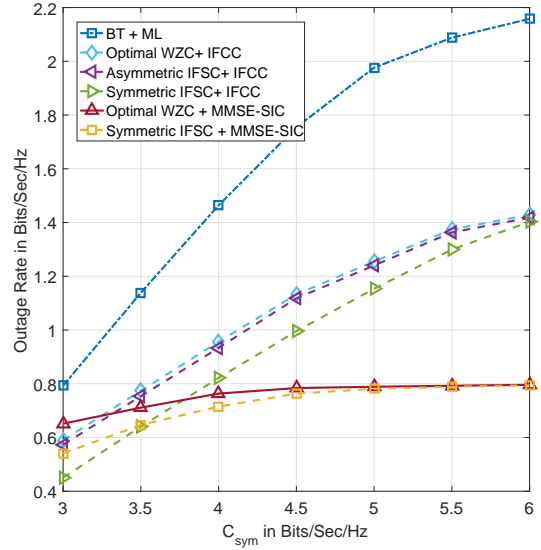}
    \caption{$L=3$ basestations.} \label{f:global_6_3}\quad\quad
  \end{subfigure}
  \vspace{-0.1in}
  \caption{$5\%$ outage rate per user with global CSIR, $K = 6$ users, and $\mathsf{SNR} = 25$dB.}\label{local_snr}
\end{figure*}

\subsection{Global CSIR}
We start by assuming global CSIR, fixing $\mathsf{SNR}=25$ dB, and plotting the outage rate per user as the fronthaul rate $\C$ varies. In Figure~\ref{f:global_3_6}, we consider a setting where there are fewer users than BSs ($K=3,L=6$), and plot the performance of various combinations of AIFSC, IFSC, WZC, and SUC source coding strategies with IFCC, MMSE-SIC, and MMSE channel decoding strategies. For WZC, we plot two variants: one that selects the optimal decompression order via exhaustive search and another that employs the heuristic decompression order from~\cite{YW16}. In this setting, the observations at the BSs are highly correlated and the source coding component plays a more important role. Indeed, there is little difference between the IFCC and MMSE-SIC versions of each architecture, and MMSE loses about $1$ bit per user. Here, WZC coupled with MMSE-SIC (both sequential decoding approaches) attains the best performance, but Symmetric IFSC coupled with IFCC follows closely behind (for which the decoding can be nearly parallelized). 

In Figure~\ref{f:convergence} also provide a sample convergence path for the symmetric distortion level as iteratively refined by  Algorithm~\ref{alg_sym_IFSC}. Recall that the asymmetric distortions are chosen in a single pass by Algorithm~\ref{AIFSC}, which seems to perform as well as iterative algorithms in our experiments. 

In Figure~\ref{f:global_6_6}, we turn to a setting where the number of users and BSs are equal ($K=L=6$). Here, source and channel coding are equally important. It is well-known~\cite{zneg14} that IFCC can significantly outperform MMSE-SIC for $K=L$, and this is confirmed by the plot, where the best performance (other than the BT+ML benchmark) is attained via IFCC. Note also that MMSE channel decoding performs quite poorly. In terms of source coding, WZC offers the best performance, but is now nearly tied with Asymmetric IFSC and closely tracked by Symmetric IFSC. These effects become more pronounced if we decrease to $L=3$ BSs. In Figure~\ref{f:global_6_3}, we see that IFCC offers a larger advantage over MMSE-SIC (and an even larger one over MMSE, which is not plotted). Again, WZC and Asymmetric IFSC are nearly equal and Symmetric IFSC can operate within a small gap.

Overall, we observe that IF source and channel coding offers strong performance as well as the possibility of parallel rather sequential decoding algorithms.

\begin{figure*}[!h]
\begin{subfigure}[t]{.49\linewidth}
    \centering\includegraphics[width=0.9\linewidth]{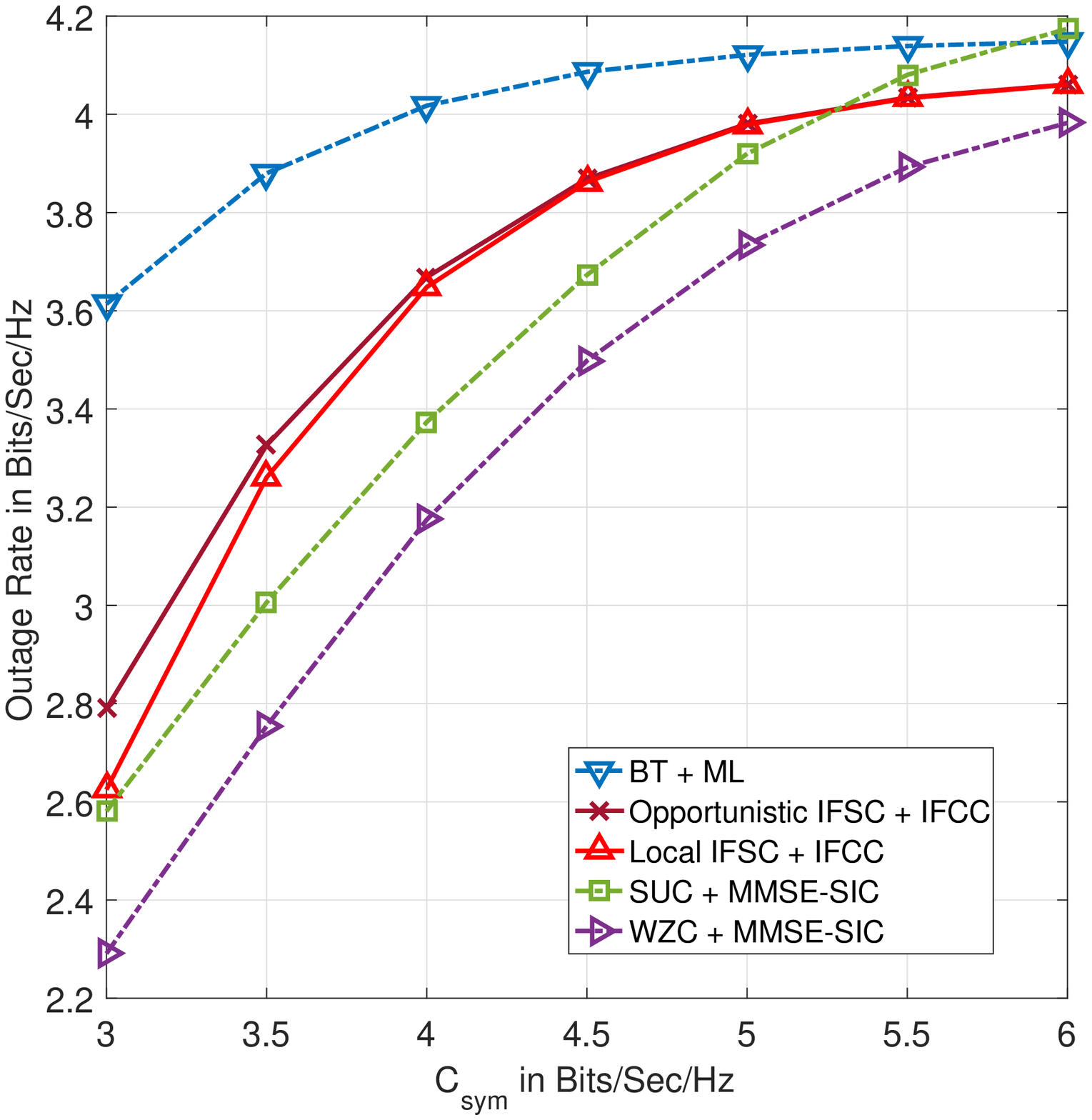}
    \caption{Fixed $\mathsf{SNR}=25$dB, varying fronthaul rate $\C$.}\label{f:local_3_6_varyC}\vspace{-0.05in}
  \end{subfigure}
  \begin{subfigure}[t]{.49\linewidth}
    \centering\includegraphics[width=0.9\linewidth]{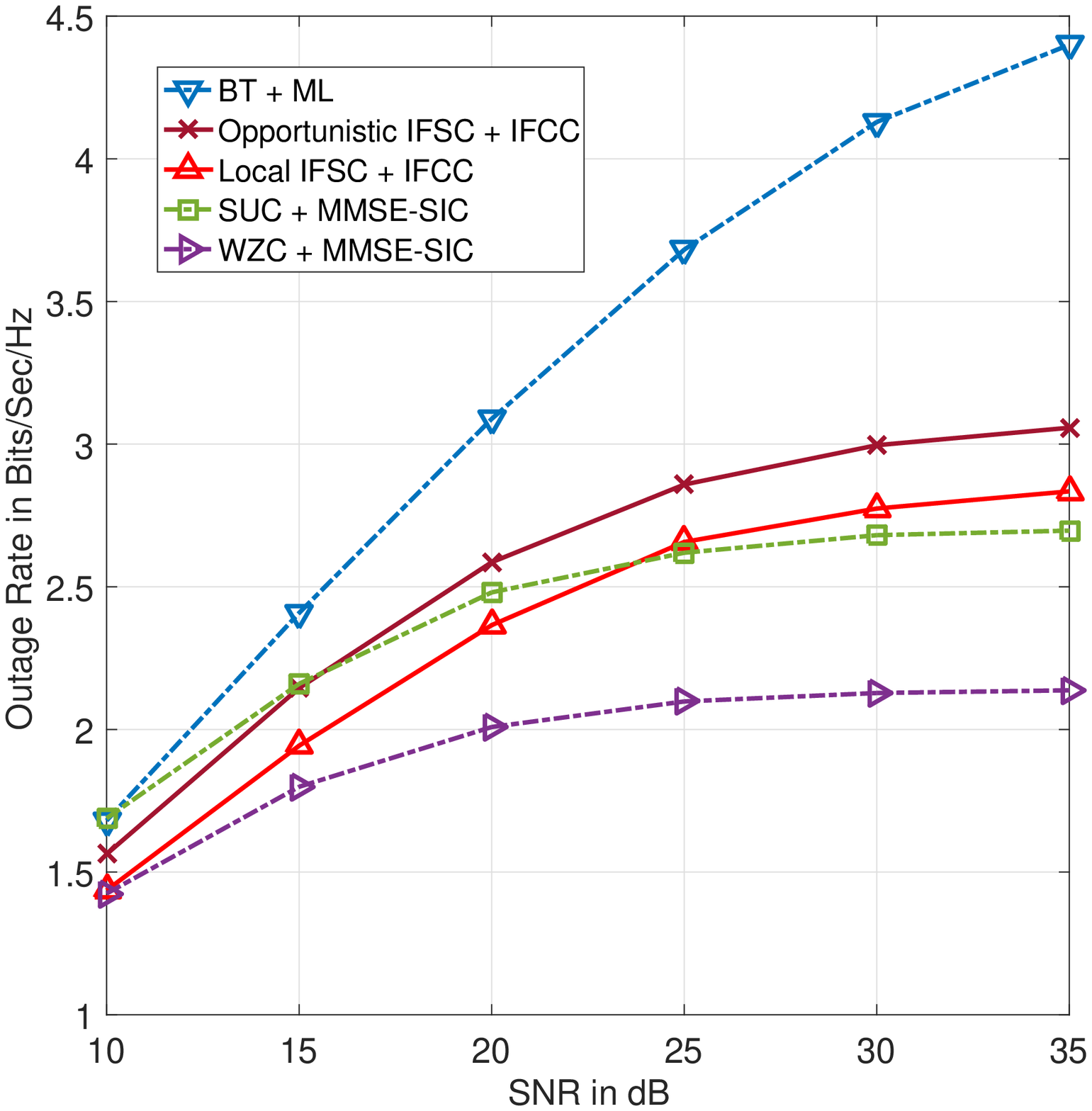}
    \caption{Fixed fronthaul rate $\C = 3$, varying $\mathsf{SNR}.$}\label{f:local_3_6_varySNR}\vspace{-0.05in}\quad
  \end{subfigure}
  \vspace{-0.01in}
  \caption{$10\%$ outage rate per user with local CSIR, $K = 3$ users, and $L = 6$ basestations.} \label{f:local_3_6}
\end{figure*} 
 
\begin{figure*}[!h]
\begin{subfigure}[t]{.49\linewidth}
    \centering\includegraphics[width=0.9\linewidth]{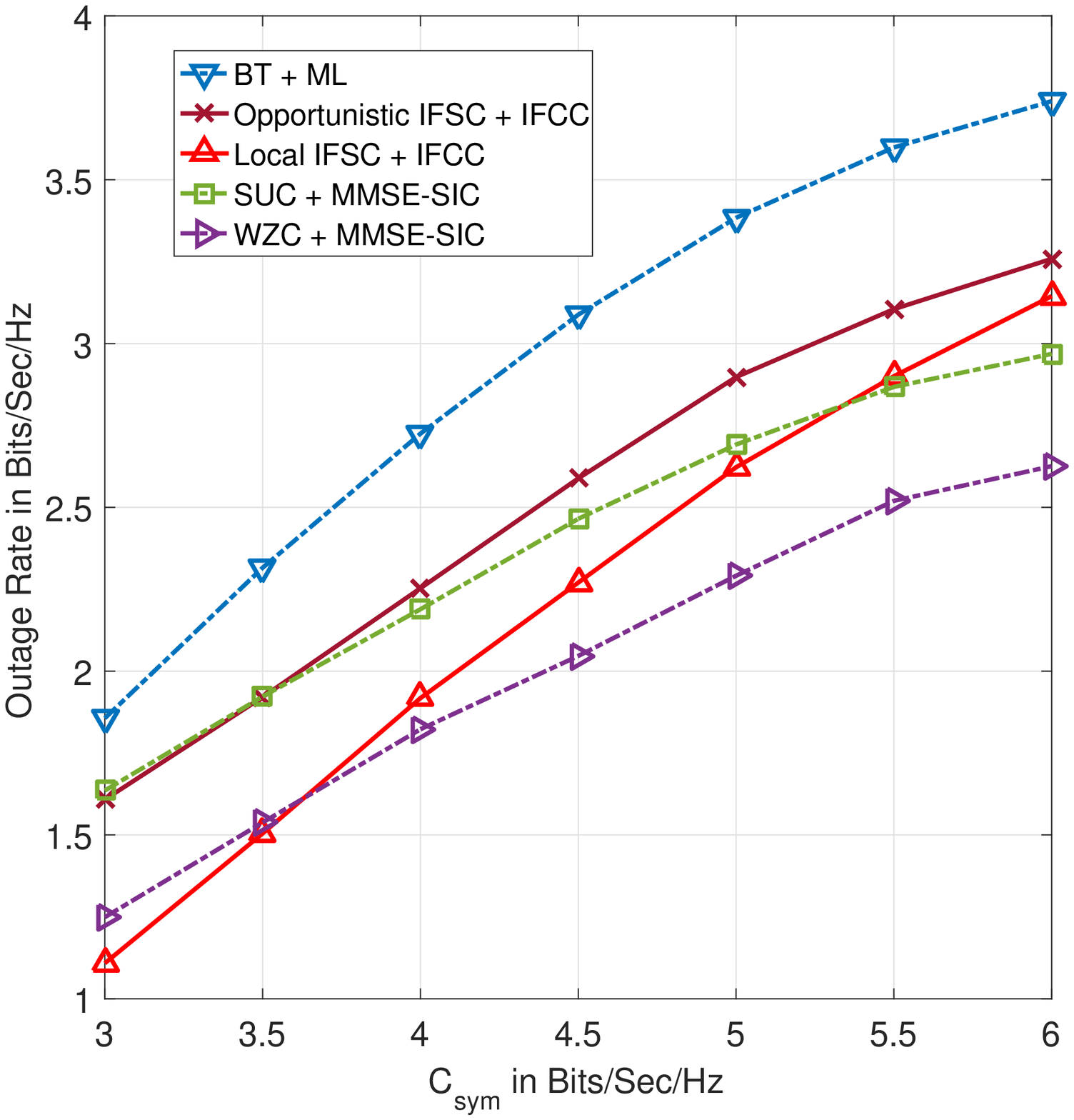}\vspace{-0.05in}
    \caption{$L=6$ basestations.}\label{f:local_6_6}
  \end{subfigure}
  \begin{subfigure}[t]{.49\linewidth}
    \centering\includegraphics[width=0.9\linewidth]{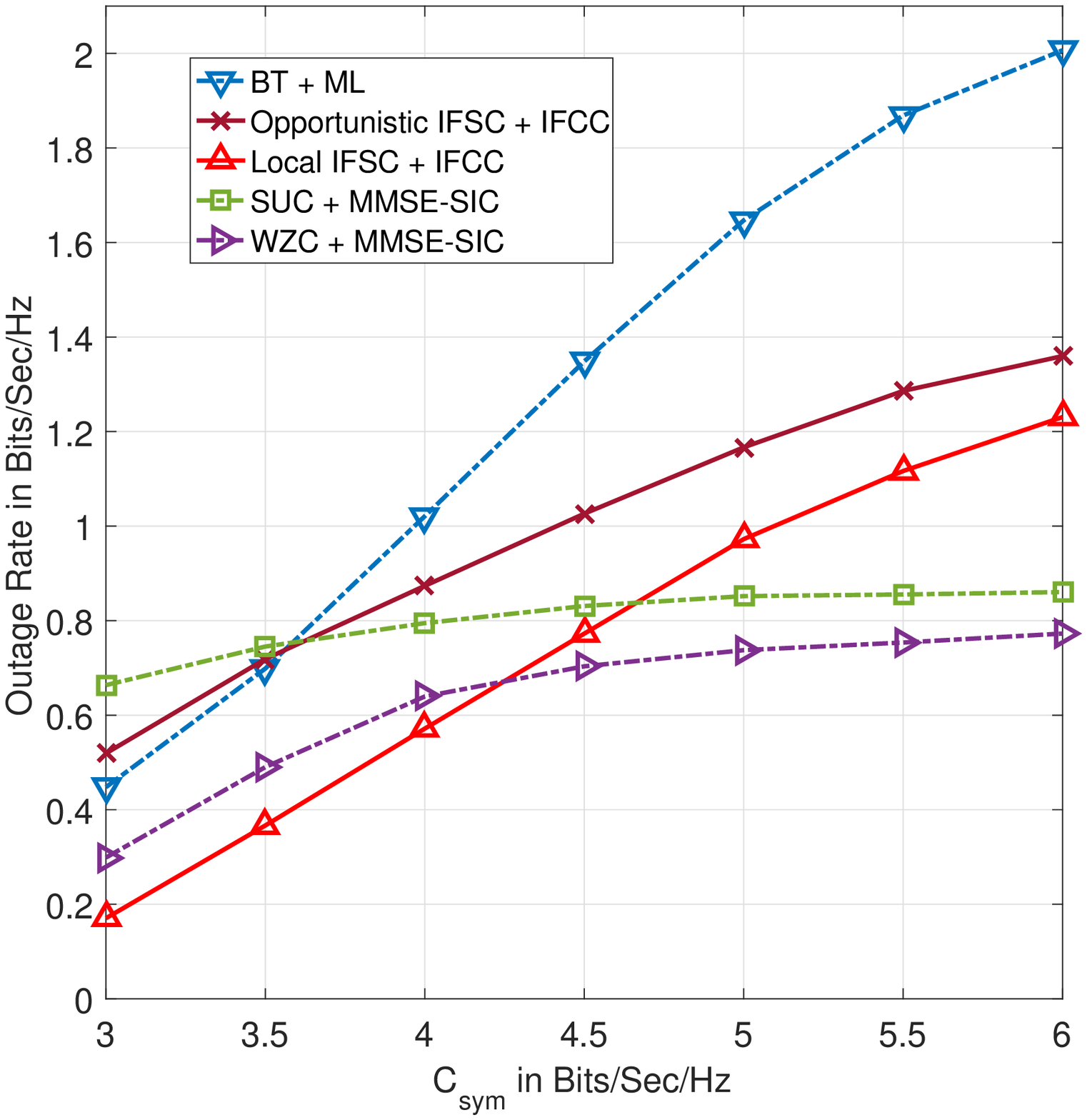}\vspace{-0.05in}
    \caption{$L=3$ basestations.}\label{f:local_6_3}\quad\quad
  \end{subfigure}
  \vspace{-0.1in}
  \caption{$10\%$ outage rate per user with local CSIR, $K = 6$ users, and $\mathsf{SNR}=25$dB.} \label{f:local_6} 
  \end{figure*}

\subsection{Local CSIR}
In the local CSIR scenario, each BS knows the channel gains to itself, and can therefore select a rate for SUC so that no outage occurs. However, to further reduce the compression rates, it must set a rate that may result in an outage, depending on the channel realizations at the other BSs. The end-to-end outage event is thus a union of the source and channel coding outage events. In these plots, we will examine the $10\%$ outage rate, allocating $5\%$ towards source coding outage and $5\%$ towards channel coding outage.

Again, we start with $K = 3$ users and $L=6$ BSs in Figure~\ref{f:local_3_6}. Recall that the the opportunistic IF strategy from Theorem~\ref{Op_IF_CRAN} switches from IFSC to SUC, if it offers a better distortion. In Figure~\ref{f:local_3_6_varyC}, we fix $\mathsf{SNR}=25$dB and vary the fronthaul rate $\C$. In this setting, opportunistic and local IFSC have essentially the same performance, which is quite close to the basic SUC strategy. Although it may seem surprising, WZC is outperformed by all three strategies, as is the case for all of our local CSIR plots. This is partly due to the fact that, since we use a symmetric distortion target for WZC, it cannot target an asymmetric corner point corresponding to an optimal sum rate. (Selecting good asymmetric distortions would require some knowledge of the channel quality order across BSs, which is not available with local CSIR.) Morever, SUC does not suffer any source coding outage, which provides an additional edge over WZC. Moving to Figure~\ref{f:local_3_6_varySNR}, we fix the fronthaul rate $\C=3$ and vary $\mathsf{SNR}$, where we observe very similar phenomena, but note that opportunistic IFSC does provide a slight advantage over local IFSC.

In Figure~\ref{f:local_6_6}, we take the number of users and BSs to be the same ($K=L=6$). The main change is that we now observe that local IFSC falls behind the performance of SUC, while opportunistic IFSC (which can select SUC when it is superior) maintains an edge. This behavior continues to hold if we lower the number of BSs to $L=3$ as seen in Figure~\ref{f:local_6_3}. Thus, even if the BSs only possess local CSIR, IF performs well across various scenarios, especially if we opportunistically mix between IFSC and SUC compression, in order to mitigate the effects of source coding outage.

\section{Conclusions}\label{sec: conclusion}
In this paper, we introduced an IF architecture for uplink C-RANs that can operate within a constant gap from the optimal tradeoff between outage rate and probability. We also proposed algorithms for efficiently selecting good integer coefficient matrices and distortion levels. We then demonstrated, via simulations, that our IF architecture is competitive with state-of-the-art C-RAN architectures based on WZC. Moreover, our IF strategy can potentially be implemented using only parallel decoding blocks, rather than the sequential decoding needed for WZC.  
 
\appendix
\subsection{Recovering $\mathbf{t}_{m,1},\ldots,\mathbf{t}_{m,L}$}\label{IFSC_SIC_sec}
The proof of the lemma below corrects a slight error in the proof of \cite[Lemma 3]{HN16}, which is needed to establish the achievable rate in Lemma \ref{lem:SIFSC} (i.e., \cite[Theorem 3]{HN16} with $\mathbf{R} = \mathbf{I}$).
\lemma\label{IFSC_SIC}
Given $\mathbf{v}_k = \sum\limits_{\ell=1}^{L} {a}_{s,k,\ell} \left( {\mathbf{y}}_\ell+\mathbf{q}_\ell \right) $ for $k=1,\ldots,m-1$ as well as $\widetilde{\mathbf{y}}_\ell=[\mathbf{y}_\ell+\mathbf{q}_\ell] \bmod \Lambda_{C,\ell}$ for $\ell=1,\ldots,L$, the CP can recover $\mathbf{t}_{m,\ell}=[\mathbf{y}_\ell+\mathbf{q}_\ell] \bmod \Lambda_{C,m}$ for $\ell=1,\ldots,L$.
	
$\textit{Proof}$ :
For $\ell \geq m$, since $\Lambda_{C,\ell} \subseteq \Lambda_{C,m}$, we can directly compute
\begin{align}
\left[ \widetilde{\mathbf{y}}_\ell \right] \bmod \Lambda_{C,m} \nonumber &=\left[ \left[ {\mathbf{y}}_\ell +\mathbf{q}_\ell \right] \bmod \Lambda_{C,\ell} \right] \bmod \Lambda_{C,m}  
\\ &\stackrel{(a)}{=} \left[ {\mathbf{y}}_\ell +\mathbf{q}_\ell \right] \bmod \Lambda_{C,m} \\ &= \mathbf{t}_{m,\ell}
\end{align}
where $(a)$ follows from the distributive law.

For $\ell<m$, we need more work to recover $\mathbf{t}_{m,\ell}$. Specifically, we cancel out the contributions of $\mathbf{t}_{m,m},\ldots,\mathbf{t}_{m,L}$ from $\mathbf{v}_1,\ldots,\mathbf{v}_{m-1}$ then solve for $\mathbf{t}_{m,1},\ldots,\mathbf{t}_{m,m-1}$. To this end, we remove the effects of the first $m-1$ dithers 
\begin{align}
\widetilde{\mathbf{v}}_k &= \mathbf{v}_k + \sum\limits_{\ell=1}^{m-1} a_{s,k,\ell} \mathbf{u}_\ell  \\ &\stackrel{(a)}{=}\sum\limits_{\ell=1}^{m-1}   a_{s,k,\ell} \left(  \mathbf{y}_\ell + \mathbf{u}_{\ell} - \left[  \mathbf{y}_\ell + \mathbf{u}_{\ell} \right] \bmod \Lambda_{F,\ell} \right) \\ & ~~~~ +~ \sum\limits_{\ell=m}^{L} a_{s,k,\ell} \left( \mathbf{y}_\ell + \mathbf{q}_\ell \right)  \nonumber \\
& \stackrel{}{=} \sum\limits_{\ell=1}^{m-1} a_{s,k,\ell} \mathcal{Q}_{\Lambda_{F,\ell}} \left( \mathbf{y}_\ell + \mathbf{u}_\ell \right) + \sum\limits_{\ell=m}^{L} a_{s,k,\ell}  \left( \mathbf{y}_\ell + \mathbf{q}_\ell \right)~~ . \label{adding_dithers}
\end{align} where $(a)$ holds since $\mathbf{q}_\ell=-[\mathbf{y}_\ell+\mathbf{u}_\ell]\bmod \Lambda_{F,\ell}$.

Now, for $k=1,\ldots,m-1$, we use $\mathbf{t}_{m,m},\ldots,\mathbf{t}_{m,L}$ to cancel out the second sum to obtain
\begin{align}
\mathbf{w}_k  &= \left[ \widetilde{\mathbf{v}}_k - \sum\limits_{\ell=m}^{L} a_{s,k,\ell} \mathbf{t}_{m,\ell} \right] \bmod \Lambda_{C,m} \nonumber \\
& \stackrel{}{=} \bigg[\sum\limits_{\ell=1}^{m-1} a_{s,k,\ell} \mathcal{Q}_{\Lambda_{F,\ell}} \left( \mathbf{y}_\ell + \mathbf{u}_\ell \right) + \sum\limits_{\ell=m}^{L} a_{s,k,\ell}  \left( \mathbf{y}_\ell + \mathbf{q}_\ell \right)\\&~~~~ -~ \sum\limits_{\ell=m}^{L} a_{s,k,\ell} \left[ \mathbf{y}_\ell +\mathbf{q}_\ell \right] \bmod \Lambda_{C,m}  \bigg] \bmod \Lambda_{C,m} \nonumber \\
& \stackrel{}{=} \left[ \sum\limits_{\ell=1}^{m-1} a_{s,k,\ell} \left[ \mathcal{Q}_{\Lambda_{F,\ell}} \left( \mathbf{y}_\ell + \mathbf{u}_\ell \right) \right]\bmod \Lambda_{C,m}  \right] \bmod \Lambda_{C,m}  \label{comb_to_be_solved}
\end{align} where the last step holds from the distributive law.

Collecting these vectors into matrices, we define $\mathbf{W}_m = [\mathbf{w}_1~\cdots~\mathbf{w}_{m-1}]^\T$,  $\widetilde{\mathbf{T}}_m = [\widetilde{\mathbf{t}}_{m,1}~\cdots~\widetilde{\mathbf{t}}_{m,m-1}]^\T$, and $\widetilde{\mathbf{t}}_{m,k} = \left[ \mathcal{Q}_{\Lambda_{F,k}} \left( \mathbf{y}_k + \mathbf{u}_k \right) \right] \bmod \Lambda_{C,m} $, so that we can write \eqref{comb_to_be_solved} as 
\begin{align}
\mathbf{W}_m = \left[ \mathbf{A}_{s, [1:m]} \widetilde{\mathbf{T}}_m \right] \bmod \Lambda_{C,m}
\end{align}

Next, we apply the inverse $\bar{\mathbf{A}}_{s,[1:m]}$ of $\left[ \mathbf{A}_{s, [1:m]} \right] \bmod p$ over $\mathbb{Z}_p$ to obtain
\begin{align*}
&\left[ \bar{\mathbf{A}}_{s,[1:m]} \mathbf{W}_m \right] \bmod \Lambda_{C,m} \\ &\stackrel{(a)}{=} \left[ \left( \left[ \bar{\mathbf{A}}_{s,[1:m]} \mathbf{A}_{s, [1:m]} \right] \bmod p   \right)\widetilde{\mathbf{T}}_m \right] \bmod \Lambda_{C,m} \\ &\stackrel{(b)}{=} \widetilde{\mathbf{T}}_m
\end{align*}
where $(a)$ holds from applying \cite[Lemma 3]{oen14} and the distributive law and $(b)$ holds from the definition of $\bar{\mathbf{A}}_{s,[1:m]}$. 
Finally, we remove the dithers we added in \eqref{adding_dithers} to obtain
\begin{align*}
&\left[ \widetilde{\mathbf{t}}_{m,\ell} - \mathbf{u}_\ell \right] \bmod \Lambda_{C,m} = \left[ \mathbf{y}_\ell + \mathbf{q}_\ell \right] \bmod \Lambda_{C,m} = \mathbf{t}_{m,\ell}
\end{align*} for $\ell=1,\ldots,m-1$. 

\subsection{Bounding $d^*$}
Recall that $\mathbf{S}_1$ is the diagonal matrix of eigenvalues from the eigenvalue decomposition of 
$\mathbf{U}\mathbf{S}_1 \mathbf{U}^\T = P \mathbf{H}^\T\mathbf{H} + \mathbf{I}.$
\lemma \label{d_bounds}
Conditioned on $\mathcal{A}$, the distortion $d^*$ satisfying $R^S_\text{IFSC}(\mathbf{H},d^*)=\C$ also satisfies $d^* > 2^{-(2\dr/L+1)}$ if $\mathbf{S}_1 \in \mathcal{B}$ and $d^* < 1$ if $\mathbf{S}_1 \in \mathcal{B}^c$ where $\mathcal{B}= \{ \frac{1}{2} \log |\mathbf{S}_1| > L\C-\dr-L/2\}$ and $\mathcal{A} = \{ R^s_\text{IFSC} (\mathbf{H}) < R^s_{\text{BT}}(\mathbf{H}) + \Delta R \}$.

$\textit{Proof}$: For $\mathbf{S}_1 \in \mathcal{B}$, assume for the sake of contradiction that $d^*\leq 2^{-(2\dr/L+1)}$, then we have 
\begin{align}
L\C -\dr -\frac{L}{2} &=LR^s_\text{IFSC}(\mathbf{H}) -\dr -\frac{L}{2} \\ & \stackrel{(a)}{\geq} L R^s_\text{BT}(\mathbf{H}) -\dr -\frac{L}{2} \nonumber \\
& = \frac{1}{2} \log \left| \frac{P}{d^*}\mathbf{H}\mathbf{H}^\T + \frac{d^*+1}{d^*}\mathbf{I} \right| -\dr -\frac{L}{2}\\ &> \frac{1}{2} \log |\mathbf{S}_1| - \frac{L}{2}\log d^* 2^{2\dr/L+1}  \\ &\stackrel{(b)}{>} \frac{1}{2} \log |\mathbf{S}_1|
\end{align} where $(a)$ holds from $R^s_\text{IFSC}(\mathbf{H}) \geq R^s_\text{BT}(\mathbf{H})$ as shown in \cite{oe17} and $(b)$ is a contradiction that holds if $d^*<2^{-(2\dr/L+1)}$.
Now, for $\mathbf{S}_1\in \mathcal{B}^c$, assume that $d^* \geq 1$ and note that 
\begin{align}
L\C -L/2 - \dr &= LR^s_\text{IFSC}(\mathbf{H})-L/2-\dr \\ & \stackrel{(a)}{\leq} L R^s_\text{BT}(\mathbf{H})-L/2 \nonumber \\ &= \frac{1}{2} \log \left| \frac{P\mathbf{H}\mathbf{H}^\T + (d^*+1)\mathbf{I}}{2d^*} \right| \\
& \stackrel{(b)}{\leq}  \frac{1}{2} \log \left| \frac{P}{2d^*}\mathbf{H}\mathbf{H}^\T + \mathbf{I} \right|  \\ &< \frac{1}{2} \log |\mathbf{S}_1|
\end{align} where $(a)$ holds from the fact that $R^s_\text{IFSC}(\mathbf{H}) \leq R^s_\text{BT}(\mathbf{H})+\dr$ and $(b)$ follows from assuming $d^*\geq 1$, which is a contradiction. Hence, we have $d^*<1$ if $\mathbf{S}_1 \in \mathcal{B}^c$.

\bibliographystyle{ieeetr}


\end{document}